\documentclass[journal ]{new-aiaa}
\usepackage{textcomp}

\usepackage{graphicx}
\usepackage{amsmath}
\usepackage[version=4]{mhchem}
\usepackage{siunitx}
\usepackage{longtable,tabularx}
\usepackage{subcaption}
\usepackage{soul}
\usepackage{algorithm}
\usepackage{algpseudocode}
\usepackage[capitalize]{cleveref}
\usepackage[percent]{overpic}

\title{Developing a Numerical Framework for the High-Fidelity Simulation of Contrails: Sensitivity Analysis for Conventional Contrails}

\author{T\^{a}nia S. C. Ferreira\footnote{Postdoctoral Fellow, Center for Turbulence Research, taniafer@stanford.edu.}, 
Juan J. Alonso\footnote{Professor, Department of Aeronautics and Astronautics, AIAA Fellow.}, and 
Catherine Gorl\'{e}\footnote{Associate Professor, Department of Civil and Environmental Engineering.}}
\affil{Stanford University, Stanford, CA 94305, USA}

\begin{document}

\maketitle

\begin{abstract}

Contrails have recently gained widespread attention due to their large and uncertain estimates of effective radiative forcing, i.e., warming effect on the planet, comparable to those of carbon dioxide.
To study this aircraft-induced cloud formation in the context of current conventional fuels and future alternative fuels, we have developed a numerical framework for simulating the jet and early vortex interaction phases of contrail formation.

Our approach consists of high-fidelity, 3D large-eddy simulations (LES) of an Eulerian-Lagrangian two-phase flow using the compressible flow solver charLES.
We perform temporal simulations of the early contrail formation phases for a single linear contrail and compare the sensitivity of the results to modeling choices and atmospheric, aircraft, and engine parameters.

Specifically, we discuss how these choices and parameters affect the number of nucleated ice crystals and estimated net radiative forcing (based on an optical depth parameterization).
Our simulations show the most significant sensitivity to the aircraft size, followed by the soot number emission index and fuel consumption.
Adding atmospheric aerosol as a precursor for future studies evidences a non-linear relation previously highlighted in the literature between number of emitted soot and nucleated ice crystals.

\end{abstract}

\section*{Nomenclature}

\noindent(Nomenclature entries should have the units identified)

{\renewcommand\arraystretch{1.0}
\noindent\begin{longtable*}{@{}l @{\quad=\quad} l@{}}
$\alpha$    & (ice) deposition coefficient \\
$b$     &   wingspan, \si{\meter} \\
$\overline{\dot \omega_{\text{v}}}$ & gas/particle mass exchange source term \\
$\rho$      & density, \si{\kilo\gram\per\cubic\meter} \\ 
$\mu$       & viscosity, \si{\kilogram\per\meter\per\second} \\
$\lambda_{\text{v}}$ & water vapor mean free path, \si{\meter} \\
$\tau_{ij}$ & stress tensor, \si{\kilogram\per\meter\per\square\second} \\
$c_{\text{p}}$       & specific heat capacity at constant pressure \\
$D_{\text{v}}$       & water vapor diffusion coefficient \\
$E$         & total energy  \\ 
$\text{EI}$         & emission index  \\ 
$e$         & internal energy  \\ 
$\eta_\text{prop}$         & overall propulsive efficiency, \%  \\ 
$g$     & gravitational acceleration \\
$G$     & collision factor \\
$\Gamma$    & (aircraft-induced) circulation \\
$k$     & thermal conductivity \\
$\text{Kn}$    & Knudsen number \\
$m_{\text{p}}$ & particle mass, \si{\kilogram} \\
$n_c$   & number of computational particles \\
$n_{\text{p}}$   & number of physical particles per \textit{n\textsubscript{c}} \\
$n_{\mathcal{V}}$   & number of particles in a control volume \\
$p$ & pressure \\
$p_{\text{v}}$   & water vapor partial pressure, \si{\pascal} \\
$\text{Pr}$ & Prandtl number \\
$r$ &   radius, \si{\meter} \\
$R$  & specific gas constant \\
$q_{j}$ & heat flux \\
$r_{\text{p}}$ & particle radius, \si{\meter} \\
$T$ &   Temperature, \si{\kelvin} \\
$\mathbf{u}$  & velocity vector, \si{\meter\per\second} \\
$\mathcal{V}$ & volume of a control volume \\
$\mathbf{x}_{\text{p}}$  & particle position vector, \si{\meter} \\
$Y_{\text{v}}$  & water vapor mass fraction \\
\end{longtable*}}
 \vspace{-2mm}
Subscript \vspace{-2mm}
{\renewcommand\arraystretch{1.0}
\noindent\begin{longtable*}{@{}l @{\quad=\quad} l@{}}
$\text{a}$             & atmospheric/ambient \\
$\text{a/c}$             & aircraft \\
$\text{core}$   & engine core \\
$\text{f}$             & fuel \\
$\text{j}$             & jet \\
$\text{i}$             & ice (crystal) \\
$\text{p}$             & particle \\
$\text{sat}$    & saturated \\
$\text{tot}$   & total: engine core and bypass \\
$\text{v}$      & water vapor \\
$v$             & vortex \\
$\text{w}$             & water (droplet) \\
\end{longtable*}}

\section{Introduction}

Aviation is a growing and hard-to-abate sector; as other transportation sectors can more easily reduce their emissions or electrify, the share of aviation-induced climate impact will take on a more significant portion.
And while the current focus is on decarbonization, non-CO\textsubscript{2} effects, such as nitrogen oxides and contrail cirrus, contribute to two-thirds of the climate impact of aviation \cite{lee2021contribution}. 
Contrail cirrus -- ice clouds formed by aircraft engines in a sufficiently cold and humid atmosphere -- are estimated to account for 57\% of aviation's climate impact, but the uncertainty on this estimate is about 70\% for conventional fuels \cite{lee2021contribution}. 
A better and more detailed understanding of the formation of these clouds is thus necessary to support large atmospheric codes that perform global radiative forcing calculations.

The lifespan of a contrail varies widely across both temporal and spatial scales, as outlined in detailed reviews  \cite{paoli2016contrail,schumann2017,karcher2018formation}.
Contrails can range from a few meters and seconds, when they begin as engine exhaust plumes forming their first ice crystals, to a kilometer-wide contrail that can persist for hours if the atmosphere has sufficient humidity with respect to ice.
The evolution of a contrail can be categorized into its formation and persistence (or spreading) stages. The characteristics of the aircraft and engine have the largest influence on the initial formation of the contrail, which can be split into the jet and vortex phases. The largest portion of the contrail's lifetime is the dissipation (also referred to as diffusion) phase, during which atmospheric variability and radiative processes determine the extent and longevity of the contrail.

During the jet phase, hot exhaust gases and particles are emitted into the atmosphere. This plume cools rapidly, and ice crystals begin to form when, and if, thermodynamic conditions permit (in short, enough humidity and freezing temperatures).
The motion of this particle-laden flow is dominated by the jet's high velocities, which subside in a matter of seconds as the jet flow loses its momentum and the wingtip vortices become the dominant flow motion.
At this stage, the contrail is in its vortex phase, in which two counter-rotating vortices descend through the stratified atmosphere. An interplay of fluid dynamics processes, such as atmospheric stratification, turbulence, and induced vortical instabilities, affects the growth and sublimation rate of the ice crystals.

Box model trajectory approaches can encapsulate largely complex chemistry and microphysics schemes to describe the evolution of ice crystals depending on prescribed atmospheric temperature and relative humidity variations but typically without resolving any fluid dynamics processes \cite{karcher1995trajectory,vancassel2014numerical,lewellen2020large,bier2022box,bier2024contrail}.
On the other hand, high-fidelity simulations can capture the intrinsic contrail physics in a very detailed manner, from the jet mixing, vortex entrainment and dissipation to stratified atmospheric turbulence, albeit with simpler microphysics models \cite{vancassel2014numerical,paoli2002contrail,paoli2004contrail,paoli2013effects,paugam2010influence,naiman2011large,picot2015large}.

Moving towards new, alternative fuels, an accurate representation of the initial conditions, the jet phase, is crucial for capturing the correct long-term properties of ice crystals.
For this reason, we focus on simulating the early phases of contrail formation, initially for conventional fuels.
Therefore, this work's first objective is to develop a numerical tool to simulate contrails with high-resolution and sufficiently detailed microphysics schemes, which will serve as a precursor for future studies of alternative fuels. 

This paper details the development of a numerical framework for simulating contrail formation and sensitivity analysis to microphysics and idealized jet modeling, as well as atmospheric, aircraft, and engine parameters.
This work focuses on the initial stages of contrail formation: the jet and early vortex interaction phases.
We organized the paper as follows: we describe the governing equations for the carrier and disperse phase along with the microphysics modeling in\cref{sec:gov_eqs}, initial and boundary conditions as well as the baseline are described in \cref{sec:IC_BC}, the two-stage simulation of contrails and sensitivity analysis are discussed in \cref{sec:sensitivity_analysis}, and conclusions are drawn in \cref{sec:conclusion}.
We further include a verification and validation analysis for the stratified vortex descent, ice deposition modeling, and jet phase of a contrail simulation, in \cref{app:VandV_strat,app:VandV_ice,app:VandV_jet}, respectively.

\section{Governing equations} \label{sec:gov_eqs}

We perform large eddy simulations (LES) of flow dynamics and ice microphysics of contrails using a two-phase flow, Eulerian-Lagrangian approach.
An Eulerian grid approach solves the compressible Navier-Stokes equations of the carrier phase (air and water vapor), and a Lagrangian particle tracking method deals with the disperse phase (particles such as aerosol, water droplets, ice crystals).

We use the charLES code from Cadence (previously Cascade Technologies), a 3D massively parallel, finite volume, compressible flow solver.
This solver uses a 2\textsuperscript{nd} order central scheme for the space discretization of the carrier phase and an explicit three-stage Runge-Kutta method for the time discretization for both carrier and disperse phases.

\subsection{Carrier phase -- Eulerian grid}

The governing equations for the low-pass filtered, compressible equations for conservation of mass, momentum, total energy and transport of water vapor are given by,

\begin{equation} 
      \frac{\partial \bar{\rho}}{\partial t} 
    + \frac{\partial (\bar{\rho} \tilde u_j) }{\partial x_j} 
    = \overline{\dot \omega_{\text{v}}},
\end{equation}

\begin{equation} 
    \frac{\partial (\bar{\rho} \tilde u_i)}{\partial t} 
    + \frac{\partial (\bar{\rho} \tilde u_j \tilde u_i ) }{\partial x_j} 
    + \frac{\partial \bar p}{\partial x_i} 
    = \frac{\partial \tilde \tau_{ij}}{\partial x_j} 
    - \frac{\partial \tau_{ij}^\text{sgs}}{\partial x_j},
\end{equation}

\begin{equation} 
    \frac{\partial ( \bar{\rho} \tilde E)}{\partial t} 
    + \frac{\partial [( \bar{\rho} \tilde E 
    + \bar p) \tilde u_j ]}{\partial x_j}
    = \frac{\partial (\tilde u_i \tilde \tau_{ij})}{\partial x_j} 
    - \frac{\partial (\tilde u_i \tau_{ij}^\text{sgs})}{\partial x_j} 
    + \frac{\partial}{\partial x_j} \left( k \frac{\partial \tilde T}{\partial x_j} \right)  
    - \frac{\partial q_{j}^\text{sgs}}{\partial x_j},
\end{equation}

\begin{equation} 
    \frac{\partial ( \bar \rho \tilde Y_{\text{v}})}{\partial t} 
    + \frac{\partial (\bar{\rho} \tilde Y_{\text{v}} \tilde u_j) }{\partial x_j} 
    = \frac{\partial}{\partial x_j} \left( D_{\text{v}} \frac{\partial \tilde Y_{\text{v}}}{\partial x_j} \right) 
    - \frac{\partial \zeta_{j}^\text{sgs}}{\partial x_j} 
    + \overline{\dot \omega_{\text{v}}},
\end{equation}
where $\rho$, $p$, $T$, $u_i$ and $Y_{\text{v}}$ refer to the fluid density, pressure, temperature, velocity vector and water vapor mass fraction.
The total energy is defined as $\tilde E = \tilde e + \tilde u_k \tilde u_k/2$ where $\bar \rho \tilde e = \bar p / (\gamma - 1)$ for a calorically perfect gas \cite{toro2009}.
The resolved stress tensor is $\tilde \tau_{ij} = 2 \mu(\tilde T) (\tilde S_{ij} - 1/3\tilde S_{kk} \delta_{ij})$ where the deviatoric stress is $\tilde S_{ij} = 1/2(\partial \tilde u_i/ \partial x_j + \partial \tilde u_j / \partial x_i)$.
The $(\bar \cdot)$ and $(\tilde \cdot)$ symbols correspond to the Reynolds and Favre averaging operators.
The source terms $\overline{\dot \omega_{\text{v}}}$, $\overline{\dot \omega_m}$, and $\overline{\dot \omega_h}$ are responsible for the coupling of water vapor mass, momentum and energy coupling in the carrier and disperse phase, as explained in the following section.

The sub-grid scale (SGS) terms, identified by the subscript $\text{sgs}$, are the SGS stress tensor $\tau_{ij}^\text{sgs}$, the SGS heat flux  $q_{j}^\text{sgs}$, and the SGS scalar flux $\zeta_{j}^\text{sgs}$, which account for the unresolved eddies in the following form,

\begin{equation}
    \tau_{ij}^\text{sgs}     = \bar \rho (\widetilde{u_i u_j}        - \tilde u_i    \tilde u_j),  \ \ \ \ \ \ \ \ 
    q_{j}^\text{sgs}        = \bar \rho (\widetilde{c_{\text{p}} T u_j} - c_{\text{p}}  \tilde T      \tilde u_j), \ \ \ \ \ \ \ \ 
    \zeta_{j}^\text{sgs}    = \bar \rho (\widetilde{Y_{\text{v}} u_j}        - \tilde Y_{\text{v}}    \tilde u_j).
\end{equation}
These terms are modeled through the sub-grid scale eddy viscosity concept as it follows,

\begin{equation}
    \tau_{ij}^\text{sgs}     = - 2 \mu_\text{sgs}(\tilde S_{ij} - 1/3\tilde S_{kk} \delta_{ij}),  \ \ \ \ \ \ \ \ 
    q_{j}^\text{sgs}        = - \frac{\mu_\text{sgs} c_{\text{p}}}{\text{Pr}_\text{t}}\frac{\partial \tilde T}{\partial x_j}, \ \ \ \ \ \ \ \ 
    \zeta_{j}^\text{sgs}    = - \frac{\mu_\text{sgs}}{\text{Sc}_\text{t}}\frac{\partial \tilde Y_{\text{v}}}{\partial x_j}, 
\end{equation}
where $\mu_\text{sgs}$ is computed through the static coefficient Vreman eddy viscosity model \cite{vreman2004eddy} and the remaining diffusion coefficients rely on the assumption of a constant turbulent Prandtl and Schmidt number, $\text{Pr}_\text{t}$ and $\text{Sc}_\text{t}$ respectively.
We are currently using the same value for both, $\text{Pr}_\text{t}=\text{Sc}_\text{t}$, which warrants a turbulent Lewis number $Le_t=1$, and ensures an equivalent turbulent diffusivity of heat and water vapor mass.

The volumetric source term for mass $\overline{\dot \omega_{\text{v}}}$ is defined a, 

\begin{equation}
\label{eq:couplingMass}
    \overline{\dot \omega_{\text{v}}}(\mathbf{x}) = - \frac{1}{\mathcal{V} (\mathbf{x})} \sum_{p=1}^{n_{\mathcal{V} (\mathbf{x})}} \frac{dm_{\text{p}}}{dt}  n_{\text{p}},
\end{equation}

where ${n_{\mathcal{V} (\mathbf{x})}}$ is the total number of computational particles inside a control volume $\mathcal{V}$ defined by the cell center position $\mathbf{x}$; $n_{\text{p}}$ is the number of physical particles each computational particle $p$ represents ($p=1, ..., n_c$, where $n_c$ i the total number of computational particles).
The modeling for the mass change $dm_{\text{p}}/dt$ according to particle microphysics is described in the following section.

\subsection{Disperse phase -- Lagrangian particle tracking} \label{sec:lagrangian}

Given the small timescales associated with momentum and heat exchanges, we assume the particles behave as a tracer for small enough Stokes numbers (our scales range from \num{e-9} to \num{e-5}). 
In neglecting momentum and heat exchanges, the single particle equations reduce to,

\begin{equation}
\label{eq:particleDx_tracer}
    \frac{d\mathbf{x}_{\text{p}}}{dt} = \mathbf{u} \left( \mathbf{x}_{\text{p}} , t \right), 
\end{equation}
in which the particle follows the flow parcel.
Assuming this equilibrium, each particle will have a velocity and temperature equal to the one of the carrier flow at their point in space (i.e., control volume) and time.

Simulating the real soot number engine emissions would require handling \num{e10} to \num{e16} computational particles inside the domain, which is altogether too costly.
However, each computational particle can represent any number of physical particles -- if we assume that one computational particle represents $n_{\text{p}}$ physical particles with the same properties (size, density, velocity, temperature, among others).
This allows us to reach a number of representative particles $n_r$ ($n_r=n_c n_{\text{p}}$) that is coherent with, e.g., an engine's soot number emission index or the atmospheric background particle concentration. 
In our sensitivity analysis, we initial ensured that the number of computational particles was above \num{e6} as \cite{paoli2013effects} reports that no sensitivity in their results for a particle number between \num{e6} and \num{e8}.
Likely due to the smaller jet diameter we use in our simulations, we neither find a distinct sensitivity between a number of \num{0.5e6} and \num{1.0e6} computational particles.

In this work, we compare two microphysics modeling methods to estimate the mass of ice crystals.
The first, assumes that particles become ice crystals as soon the surrounding water vapor is saturated with respect to water (neglecting condensation and freezing processes) and models their growth with a diffusion growth model.
This approach has been widely used in LES studies \cite{paoli2002contrail,paoli2004contrail,paoli2013effects,paugam2010influence,naiman2011large,picot2015large}.
The second approach is more comprehensive; it models the aerosol activation into water droplets, condensation growth, homogeneous freezing, and deposition growth (including latent heat effects).
It has been used in box trajectory models and some LES studies \cite{lewellen2020large,bier2022box,bier2023contrail}.

\subsubsection*{Ice microphysics model} \label{sec:ice_model}

\begin{figure}[hbt!]
\centering
\includegraphics[width=.4\textwidth]{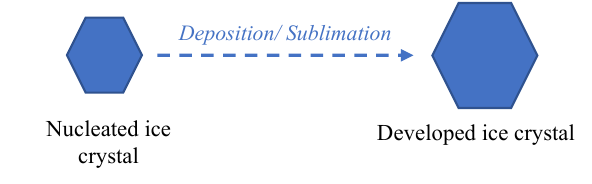}
\caption{Sketch of ice microphysics modeling.}
\label{fig:iceCrystals}
\end{figure}

Our initial approach for the microphysics of ice crystals is starting with nucleated ice crystals of a certain size and modeling their mass gain or loss by deposition or sublimation, respectively, as illustrated in \cref{fig:iceCrystals}.
The details of this modeling and other physical proprieties are described below.

The deposition model for ice crystal growth is based on the diffusion law by \cite{karcher1996initial}.
Several researchers, \cite{paoli2002contrail,paoli2004contrail,paoli2013effects,paugam2010influence,naiman2011large,picot2015large}, among others, have used this model for highly-resolved simulations of conventional contrails.

In our simulations, we activate this model once the vapor surrounding a particle is saturated with respect to water.
In practice, every particle has a saturation flag set to zero at the start of the simulation.
Once the partial pressure of water vapor in a control volume reaches that of saturation with respect to water, $p_{\text{v}}(\mathbf{x}_{\text{p}}) \geq p_{\text{v}_{\text{sat},{\text{w}}}}(\mathbf{x}_{\text{p}})$, the saturation flag of the particles contained in that control volume switches to one.
The deposition model is then activated, and, as long as vapor is supersaturated, now, with respect to ice, the particle representing an ice crystal will grow by deposition (uptake of water vapor) until the water vapor depletes.
In subsaturation, active particles will sublimate up to a minimum core radius. In reaching the minimum core radius ($r_{\text{p}} < r_{p_{min}}$), the saturation flag becomes zero again, deactivating the particle growth. 
This minimum core radius $r_{p_{min}}$ may represent the soot particle radius, and particles are allowed to re-nucleate and form ice crystals again as long as vapor saturation with respect to water is reached again.

In this approach, we assume particles to be spherical with a radius $r_{\text{p}}$,

\begin{equation}
    m_{\text{p}} = \frac{4}{3} \pi \rho_{\text{p}} r_{\text{p}}^3,
\end{equation}
where $\rho_{\text{p}}$ is the density of the particle. Neglecting the soot core, we assume a constant particle density of $\rho_{i}=917$ \si[per-mode=symbol]{\kilogram\per\cubic\meter}.

For each particle, the mass change is a function of the change in the radius based on the deposition model by K\"{a}rcher et al. \cite{karcher1996initial},

\begin{equation}
    \frac{dr_{\text{p}}}{dt} = \frac{G(r_{\text{p}}) D_{\text{v}}(\mathbf{x}_{\text{p}}) [\rho Y_{\text{v}} (\mathbf{x_{\text{p}}}) - \rho Y_{\text{v}_{\text{sat},{\text{i}}}}(\mathbf{x_{\text{p}}})] }{\rho_{\text{p}} r_{\text{p}}},
\end{equation}
and we use the equation of state for the saturated water vapor density, considering the compressibility factor close to one for the range of temperatures relevant for contrail formation \cite{pruppacher2010microphysics},

\begin{equation}
    \rho Y_{\text{v}_{\text{sat},{\text{i}}}}(\mathbf{x_{\text{p}}}) = \frac{p_{\text{v}_{\text{sat},{\text{i}}}}(\mathbf{x_{\text{p}}})}{R_{\text{v}} T(\mathbf{x_{\text{p}}})},
\end{equation}
where $R_{\text{v}}=461.52$ \si{\joule\per\kelvin\per\kilo\gram} is the specific gas constant for water vapor.

The collision factor $G$ accounts for the transition from the gas kinetic to the continuum regime as $G \rightarrow 1$ for $\text{Kn} \rightarrow 0$ (diffusion limit, maximum deposition rate) and  $G \rightarrow 0$ for $\text{Kn} \geq 1$ (free molecular regime, the larger the $\text{Kn}$, the slower the deposition takes place). 
The collision factor $G$ is written as, 

\begin{equation}
    G(r_{\text{p}}) = \left( \frac{1}{1 + \text{Kn}} + \frac{4}{3} \frac{\text{Kn}}{\alpha} \right) ^{-1},
\end{equation}
where $\text{Kn}$ is the the Knudsen number ($\text{Kn}=\lambda_{\text{v}} / r_{\text{p}}$), and $\alpha$ is the deposition coefficient.
In \cite{pruppacher2010microphysics}, values of $\alpha$ are reported between 0.014 and 1.0, and previous researchers have used values of $\alpha$ equal to 0.1, 0.5 or 1.0 in their simulations. This coefficient is directly proportional to the rate of ice deposition, and may, consequently, alter the number of estimated ice crystals as, e.g., a high rate of deposition quickly depletes the surrounding water vapor.

The water vapor mean free path $\lambda_{\text{v}}$ and diffusion coefficient of water vapor molecules in air $D_{\text{v}}$ depend on the temperature and pressure inside the control volume as follows \cite{pruppacher2010microphysics},

\begin{equation}
    \lambda_{\text{v}} = 6.15 \cdot 10^{-8} \left( \frac{101325}{p} \right) \left( \frac{T}{273.15} \right),
\end{equation}

\begin{equation}
    D_{\text{v}} = 2.11 \cdot 10^{-5} \left( \frac{101325}{p} \right) \left( \frac{T}{273.15} \right) ^{1.94},
\end{equation}
where  $\lambda_{\text{v}}$  is given in \si{\meter} and $D_{\text{v}}$ in \si{\meter\square\per\second} for a range of 233 < $T$ < 313 \si{\kelvin}.

\subsubsection*{Droplet activation to ice deposition microphysics model} \label{sec:full_micro_model}

\begin{figure}[hbt!]
\centering
\includegraphics[width=.6\textwidth]{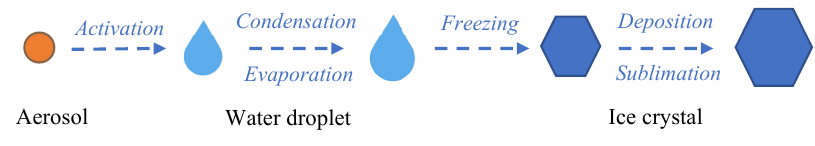}
\caption{Sketch of aerosol-to-ice microphysics modeling.}
\label{fig:fullMicro}
\end{figure}

In this paper, we adopt the approach of \cite{bier2022box,bier2024contrail} for modeling the pathway of a particle into a water droplet, and then, ice crystal, as sketched in \cref{fig:fullMicro}.
We refer to the above papers as sources of this modeling and provide a brief description of the approach below.

The model calculates the equilibrium saturation pressure of a water droplet or ice crystal surface based on the Kappa-K\"{o}hler equation from \cite{petters2007single}.
Aerosol particles have thus to overcome the Kelvin effect (surface curvature barrier) aided by the Raoul effect (solute), where the water activity is defined based on a hygroscopicity parameter $\kappa$ specific for each type of aerosol.
Both the condensational growth of activated aerosol and water droplets and the depositional growth of ice crystals include heat and mass diffusion terms, thus also accounting for latent heat.
Although the current single-particle equations assume thermal equilibrium, thereby neglecting latent heat in the gas-particle exchange, and neglect drag effects, future work will address the relevance of including this effect in the present modeling.
Droplet freezing is modeled through the homogeneous freezing temperature as defined in  \cite{karcher2015microphysical} -- as soon as the surrounding particle temperature drops below this temperature, the particle is assumed to be frozen into an ice crystal.
We assume a constant density for the droplets and ice crystals of 997 and 917 \si{\kilo\gram\cubic\meter}, respectively.

\subsubsection*{Saturation pressures}

We use the saturation pressures from \cite{murphy2005review}, where the saturation vapor pressure over liquid water is valid for a 123 < $T$ < 332 \si{\kelvin},
\begin{multline}
    \ln (p_{\text{v}_{\text{sat},{\text{w}}}}) = 54.842763 - \frac{6763.22}{T} - 4.210 \ \ln(T) + 0.000367 \ T + \tanh \{0.0415 (T - 218.8) \} \\
    \left( 53.878 - \frac{1331.22}{T} - 9.44523 \ln(T)  + 0.014025 \ T \right),
\end{multline}
and the saturation vapor pressure over ice for $T$ > 110 \si{\kelvin},

\begin{equation}
    \ln (p_{\text{v}_{\text{sat},{\text{i}}}}) = \left( 9.550426 - \frac{5723.265}{T} + 3.53068 \ln(T) - 0.00728332 \ T \right),
\end{equation}
both $p_{\text{v}_{\text{sat},{\text{w}}}}$ and $p_{\text{v}_{\text{sat},{\text{i}}}}$ are given in \si{\pascal} and depend on the temperature within the control volume containing the particle(s).

\section{Initial and boundary conditions} \label{sec:IC_BC}

The evolution of contrails can be split into four main regimes: jet, vortex, dissipation, and diffusion regimes, as described in \cite{paoli2016contrail}.
The usual procedure in previous literature separates these regimes and resorts to, e.g., idealized jets and vortices inserted at certain times in the simulation to represent the jet and vortex phases, respectively. 
Stratification becomes increasingly relevant for the vortex phase and as simulations extend to the dissipation and diffusion regimes. 

As a starting point, we follow this method to represent the dominating physics of the initial contrail phases, which are more relevant to the formation of contrails, the jet, and the vortex phases.
Below, we describe the equations we use to initialize the flow field.
We defined the boundary conditions for the contrail simulations as periodic in the contrail/jet/vortex direction and non-reflective outlet for the other directions unless otherwise stated.

\paragraph*{Idealized jet:} Jets are initialized with a hyperbolic tangent function, as described in \cite{paoli2004contrail}. The axial velocity $u_x$, temperature $T$, and water vapor mass fraction $Y_{\text{v}}$ are a function of the radial distance to the jet centerline $r$ as,

\begin{equation}
\label{eq:jet_ux}
    u_x(r) = \frac{1}{2} \left[ (u_{\text{j}} + u_{\text{a}}) - (u_{\text{j}} - u_{\text{a}}) F(r) \right],
\end{equation}

\begin{equation}
\label{eq:jet_T}
    T(r) = \frac{1}{2} \left[ (T_{\text{j}} + T_{\text{a}}) - (T_{\text{j}} - T_{\text{a}}) F(r) \right],
\end{equation}

\begin{equation}
\label{eq:jet_Y}
    Y_{\text{v}}(r) = \frac{1}{2} \left[ (Y_{\text{v}_{\text{j}}} + Y_{\text{v}_{\text{a}}}) - (Y_{\text{v}_{\text{j}}} - Y_{\text{v}_{\text{a}}}) F(r) \right],
\end{equation}

\begin{equation}
\label{eq:jet_tanh}
F(r) = \tanh \left[ \frac{1}{4} \frac{r_{\text{j}}}{\theta}  \left( \frac{r}{r_{\text{j}}} - \frac{r_{\text{j}}}{r} \right) \right],
\end{equation}
where the subscripts $\text{j}$ and $\text{a}$ refer to the jet and atmosphere, respectively.
For the baseline (narrowbody) case, a value of 10 is assigned to the ratio $r_{\text{j}} / \theta$ as in previous work \cite{paoli2004contrail}. For the aircraft size comparison, this ratio had to be changed to 30 for the widebody case to obtain a sharper jet profile when adding the bypass flow.
A random perturbation $\delta u_x$ of 1\% is usually added to the base flow velocity  $u_x$ to induce the jet's "transition" to turbulent.

To account for the colder bypass flow, in addition to the core flow, we have superimposed axial velocity and temperature profiles (water vapor mass fraction remains the same for the bypass as the atmospheric one) corresponding to the total (core and bypass) radius. 

\paragraph*{Idealized vortex:}  We add vortices as idealized vortices at a time in which the jets are fully-developed. 
We adopted the idealized Lamb-Oseen vortex equations with the following tangential velocity $u_{\theta}$ and radial pressure variation $dp/dr$,

\begin{equation}
\label{eq:LO_vortex_v_tan}
    u_{\theta}(r) = \frac{\Gamma_0}{2 \pi r} \left[ 1 - e^{ - \frac{r^2}{r_v^2} }  \right],
\end{equation}

\begin{equation}
\label{eq:LO_vortex_dp}
    \frac{dp}{dr} = \rho \frac{u_{\theta}^2}{r},
\end{equation}
where $\Gamma_0$ is the initial vortex circulation in \si{\meter\square\per\second}, and $r_v$ is the vortex core radius in \si{\meter}.

\paragraph*{Stratification:} Atmospheric stratification is set by an initialized pressure $p$, temperature $T$, and density $\rho$ field which depends on the altitude $z$.
These fields are computed from the equations written by \cite{paoli2013effects}, based on the hydrostatic balance equation, the energy conservation equation, and the perfect gas law,

\begin{equation}
\label{eq:T_strat}
    \frac{T(z)}{T_\text{ref}} = 
    \left[ \frac{g^2}{N_\text{BV}^2  c_{\text{p}}  T_\text{ref}} + 
    \left( 1 - \frac{g^2}{N_\text{BV}^2  c_{\text{p}}  T_\text{ref}} \right) 
    e^{\frac{N_\text{BV}^2}{g} (z - z_\text{ref}) } \right],
\end{equation}

\begin{equation}
\label{eq:p_strat}
    \frac{p(z)}{p_\text{ref}} = 
    \left[ \frac{g^2}{N_\text{BV}^2  c_{\text{p}}  T_\text{ref}} + 
    \left( 1 - \frac{g^2}{N_\text{BV}^2  c_{\text{p}}  T_\text{ref}} \right) 
    e^{\frac{N_\text{BV}^2}{g} (z - z_\text{ref}) } \right] 
    ^{\frac{\gamma}{\gamma-1}} e^{-\frac{\gamma}{\gamma-1}  \frac{N_\text{BV}^2}{g} (z - z_\text{ref})},
\end{equation}

\begin{equation}
\label{eq:rho_strat}
     \frac{\rho(z)}{\rho_\text{ref}} = 
    \left[ \frac{g^2}{N_\text{BV}^2  c_{\text{p}}  T_\text{ref}} + 
    \left( 1 - \frac{g^2}{N_\text{BV}^2  c_{\text{p}}  T_\text{ref}} \right) 
    e^{\frac{N_\text{BV}^2}{g} (z - z_\text{ref}) } \right] 
    ^{\frac{1}{\gamma-1}} e^{-\frac{\gamma}{\gamma-1}  \frac{N_\text{BV}^2}{g} (z - z_\text{ref})},
\end{equation}
where $N_\text{BV}$ is the Brunt-V\"{a}is\"{a}l\"{a} frequency in \si{\per\second} and $T_\text{ref}$ is the temperature at the reference altitude $z_\text{ref}$. 
We assume the following values: the gravitational acceleration as $g=9.81$ \si{\meter\per\square\second}, the specific heat capacity at constant pressure as $c_{\text{p}}=1004.8$ \si{\joule\per\kilogram\per\kelvin}, and the specific heat ratio as $\gamma=1.4$.

In addition to the above initial conditions, we also need to add a source term in the vertical momentum and in the energy conservation equation,

\begin{equation}
\label{eq:rho_uz_strat}
\overline{\dot S_{\rho u_z}} = \rho g,
\end{equation}

\begin{equation}
\label{eq:rho_e_strat}
\overline{\dot S_{\rho e}} = \rho u_z g,
\end{equation}
where $u_z$ is the velocity in the vertical direction.

\paragraph*{Baseline:}

All the cases are based on a common narrowbody two-engine aircraft (akin to an A320/B737), except for the comparison with a widebody representing a large two-engine B777/A350 aircraft in \cref{sec:sensitivity_ac_engine}.
The atmospheric parameters are the similar to the ones reported in Paoli et al. \cite{paoli2013effects} except for a lower relative humidity over ice which is 110\% in the current baseline.
The engine conditions are based on the values used for previous engine exhaust simulations \cite{garnier1997engine}.
More details can be found in \cref{tab:atm_jet_base,tab:aircraft_narrowbody}.

\begin{table}[hbt!] \centering
\caption{Atmospheric and jet exhaust properties.} \label{tab:atm_jet_base} 
\begin{tabular}{l|cccc} \hline
Properties & $T$ {[}\si{\kelvin}{]}  & $u_x$ {[}\si{\meter\per\second}{]} & $Y_{\text{v}}$ \\ \hline
Atmosphere   & 220   & 0       & \num{7.56e-5} \\
Bypass jet   & 242   & 311.6   & \num{7.56e-5} \\
Core jet     & 580   & 480.3   & \num{3.96e-2} \\ \hline
\end{tabular}
\end{table}

The total number of soot particles emitted along the simulation domain is derived based on the number of soot particles per meter of plume $N_{\text{soot}}=\text{EI}_{\text{soot}} \dot{m}_{\text{p}} / V_{\text{a/c}} = 1.3 \cdot 10^{11}$ \si{\per\meter}.
This corresponds to a total of $N_{\text{p}}$ \num{4.9e11} particles for the axial length of $12r_{\text{j}}$.
We use about \num{0.5e6} computational particles multiplied by \num{8.6e5} representative particles to achieve the $N_{\text{p}}$ value for the baseline case.
Particles are initialized with a random location at the inner core of the jet and a monodisperse radius of 20 \si{\nano\meter}.
The latter is a reasonable assumption as the past comparison between a constant and log-normally distributed particle diameter did not evidence any difference in their final size distribution \cite{paoli2013effects}.

\begin{table}[hbt!] \centering
\caption{Properties of a representative narrowbody aircraft and engine.}
\label{tab:aircraft_narrowbody} 
\begin{tabular}{llll|llll}  \hline
\multicolumn{4}{c|}{Narrowbody aircraft}                       & \multicolumn{4}{c}{Engine}               \\  \hline
Flight velocity  & $V_{\text{a/c}}$ & 252  & \si{\meter\per\second}    & Fuel mass flow    &   $\dot{m}_{\text{f}}$            & 0.34  & \si{\kilogram\per\second}  \\
Circulation  & $\Gamma$      & 290    & \si{\meter\squared\per\second} & Soot emission index &  $\text{EI}_{\text{soot}}$   &  \num{e14}  & \si{\per\kilogram} fuel         \\
Wingspan             & $b$            & 35.7 & \si{\meter}              & Water  emission index    &   $\text{EI}_{\text{H$_2$0}}$ & 1.25   & \si{\kilogram\per\kilogram} fuel \\
Vortex core distance & $b_0$          & 28   & \si{\meter}              & Water mass emission     &   $m_{\text{Y}}$               & 0.9 & \si{\gram\per\meter} \\
Vortex core radius   & $r_v$          &  2   & \si{\meter}              & Jet core radius    &  $r_{\text{core}}$  &    0.305   &\si{\meter}         \\
Jet/vortex distance  & $d_{\text{j/v}}$ & 8   & \si{\meter}             & Jet total radius &  $r_{\text{total}}$   &   0.755  & \si{\meter}         \\\hline
\end{tabular}
\end{table}

The simulations are run in two stages: the jet (decay) and early vortex (jet-vortex interaction) phase. The jet phase requires higher resolution to resolve the initially small plume; the next stage, adding the vortex, requires a larger area of refinement. To comply to both needs without incurring in prohibitive computational costs, two meshes with the same domain but different refinements are used.

The boundary conditions are periodicity in the axial direction of the jet and non-reflective outlet constant pressure for the remaining directions.
We assume a constant turbulent Prandtl and Schmidt number where $\text{Pr}_\text{t}=\text{Sc}_\text{t}=$ 0.419.
For the ice deposition modeling, we use a deposition coefficient $\alpha$ of 0.5 to be equal to its equivalent, the mass accommodation coefficient for deposition $\alpha_{\text{m,d}}$, reported in \cite{bier2022box} for the aerosol-to-ice microphysics modeling.
The hygroscopicity $\kappa$ is assumed equal to 0.005 for coated aircraft soot particles as in Ref.~\cite{karcher2015microphysical,lewellen2020large,bier2022box,bier2024contrail}.

The mesh size is $L_x = 12 r_{\text{j}}$ \si{\meter}, $L_y = 600$ \si{\meter}, and $L_z = 900$ \si{\meter} to allow enough vertical height as the vortex descends.
Mesh is refined for 16 points across the jet for the initial jet phase simulations, and for 20 points across the vortex in jet-vortex interaction simulations.
This refinement region encompasses a box of size $ 12 \times 56 \times 16 $ $r_{\text{j}}^3$ \si{\cubic\meter} for the jet phase and of size $ 12 \times 260 \times 100$ $r_{\text{j}}^3$ \si{\cubic\meter}  for the jet-vortex interaction.

\section{Comparing contrail effects}

In this work, we mainly analyze contrail formation in terms of the number of nucleated ice crystals and estimated net radiative forcing.
The number of ice crystals helps discern the individual contribution of soot particles and ambient aerosol to the formation of contrails.
The net radiative forcing provides an easy single measure to compare the warming potential of these contrails. 
This radiative forcing, which is based on a parameterization of the optical depth,  is averaged in the flight direction and integrated over the contrail width.
Below, we briefly describe how we compute the optical depth and radiative forcing parameters.

\subsection{Optical depth}

The optical depth $\tau(x,y)$ is first computed vertically across the contrail, resulting in a 2D top view of the contrail optical thickness.
The particles representing ice crystals are binned into vertical columns of size $\Delta x \times \Delta y$, as described by \cite{paoli2013effects}. 
Their optical depth is then integrated over the vertical direction, which, considering $N_{\textrm{vert}}$ computational particles inside the column, each one representing $n_{\text{p}}$ physical particles, leads to

\begin{equation}
\tau(x,y) = \sum_{\text{p}=1}^{N_{\textrm{vert}}} \frac{\pi r_{\text{p}}^2}{\Delta x \, \Delta y} Q_{\textrm{ext}}(r_{\text{p}}) n_{\text{p}},  
\end{equation}
where $r_{\text{p}}$ is the ice crystal radius. 
The Mie extinction coefficient $Q_{\textrm{ext}}$ at a given wavelength $\lambda$ is approximated by \cite{hulst1981light} and is valid for refractive indices of $1> \mu > 2$, 

\begin{equation}
Q_{\textrm{ext}} = 2 - \frac{4}{\rho} \left[ \sin(\rho) - \frac{1-\cos(\rho)}{\rho}\right], \quad \rho = \frac{4\pi r_p(\mu-1)}{\lambda}.
\end{equation}
We assume $\mu=$ 1.31 and $\lambda=0.55$ \si{\micro\meter} for ice crystals as in \cite{karcher2009aerodynamic}.

We average the optical depth $\tau(x,y)$ in the streamwise (flight) direction to obtain the optical depth along the contrail width $\tau(y)$ and, finally, compute the ninetieth percentile $\tau_{90}$ to compare time histories of optical depth across contrails.

\subsection{Radiative forcing}

The net radiative forcing (RF) is estimated based on a parameterization of the optical depth described in \cite{sanz2021impacts} for a cloud-contrail overlap study and originally developed for clouds by \cite{corti2009simple}.
The net radiative forcing represents a balance between the short-wave radiation forcing $\text{RF}_{\text{SW}}$ and the long-wave radiation forcing $\text{RF}_{\text{LW}}$, simply,
\begin{equation}
    \text{RF} = \text{RF}_{\text{SW}} + \text{RF}_{\text{LW}}.
\end{equation}
For the sake of briefness, we only include the short form of the radiative forcing definitions and the reader is referred to the above references for more detailed information.
The $\text{RF}_{\text{SW}}$ is summarized as,
\begin{equation}
    \text{RF}_{\text{SW}} = - S \cdot t \cdot (1 - \alpha) \frac{R_{\text{c}}(\tau) - \alpha R_{\text{c}}'(\tau)}{1 - \alpha R_{\text{c}}'(\tau)},
\end{equation}

where $\tau$ is the optical depth, $\alpha$ is the surface albedo, $S=S(\theta)$ is the solar flux, $\theta$ is the solar zenith angle, and $J$ the Julian day taken as 152 (June 1$^{\mathrm{st}}$).

The $\text{RF}_{\text{LW}}$ is defined as
\begin{equation}
    \text{RF}_{\text{LW}} = \epsilon(\tau) \times \text{OLR}_\text{clear} - \epsilon(\tau) \sigma^* T_c^{k*},
\end{equation}
where $\epsilon$ is the contrail emissivity, $\text{OLR}_\text{clear}$ is the longwave radiation from the Earth's surface and $\sigma^*$ (adjusted Stefan-Boltzmann constant) and $k*$ are constants as defined in \cite{corti2009simple}.

\section{Results} \label{sec:sensitivity_analysis}

First, we describe the simulation steps of our new platform for high-fidelity simulations of the first stages of contrail formation.
These are split in the initial jet phase and the jet-vortex interaction phase, in \cref{sec:results_jet_phase,sec:results_jet_vort_interaction}, respectively, for a single jet and vortex with a symmetry boundary condition, i.e., a half domain simulation.

For the sensitivity analysis, we simulate the full domain with an interacting jet and vortex pair. 
In \cref{sec:sensitivity_meshing}, we discuss how sensitive our simulations are to the meshing, number of computation particles, and domain size.
We look into the impact of a more detailed contrail modeling by adding latent heat, an idealized co-axial jet to represent the bypass flow, and by extending the microphysics modeling to include water droplet activation, condensation, and freezing in \cref{sec:sensitivity_modeling}.
Finally, we analyze the effect of changes in atmospheric temperature and relative humidity, aircraft size and fuel consumption, and soot particle number emissions in \cref{sec:sensitivity_ac_engine}.

\subsection{Initial jet phase} \label{sec:results_jet_phase}

As in previous LES studies on the early contrail formation \cite{gago2002numerical,paoli2003dynamics,paoli2004contrail,paoli2013effects,picot2015large}, simulations are split into two stages: the initial jet phase and the ensuing vortex phase. 
This is in consequence of the differing mesh resolution requirements.
The jet phase is modeled as an idealized jet with 1\% random perturbations in the axial velocity. As such, it will transition from a uniform axial momentum to a turbulent decaying jet in the atmosphere.

\begin{figure*}[hbt!] \centering
    \begin{subfigure}[t]{0.30\textwidth} \centering
        \begin{overpic}[width=\textwidth]{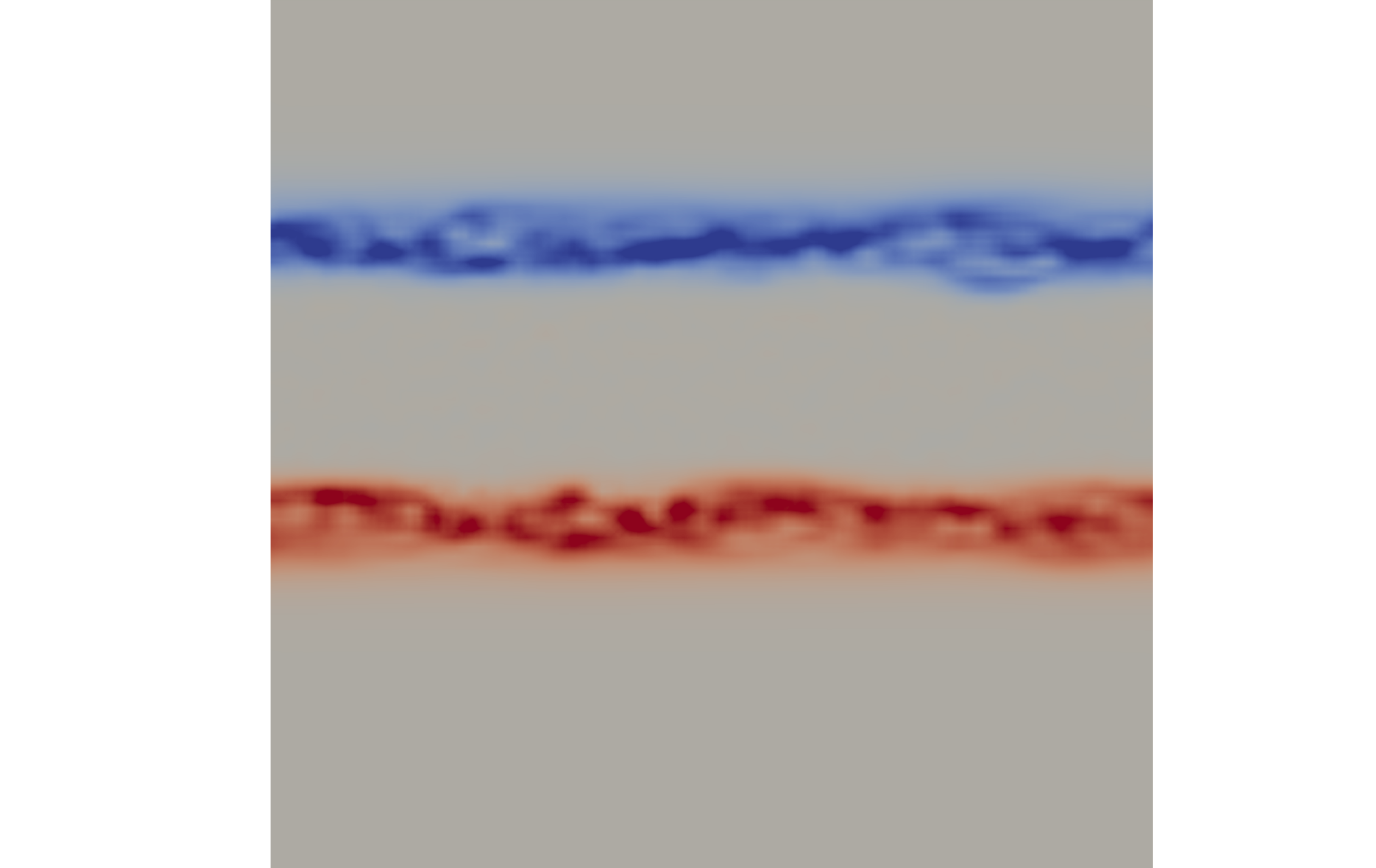}
            \put(30,60){}{}{(a)}
        \end{overpic}
    \end{subfigure}
    \hfill
    \begin{subfigure}[t]{0.30\textwidth} \centering
        \begin{overpic}[width=\textwidth]{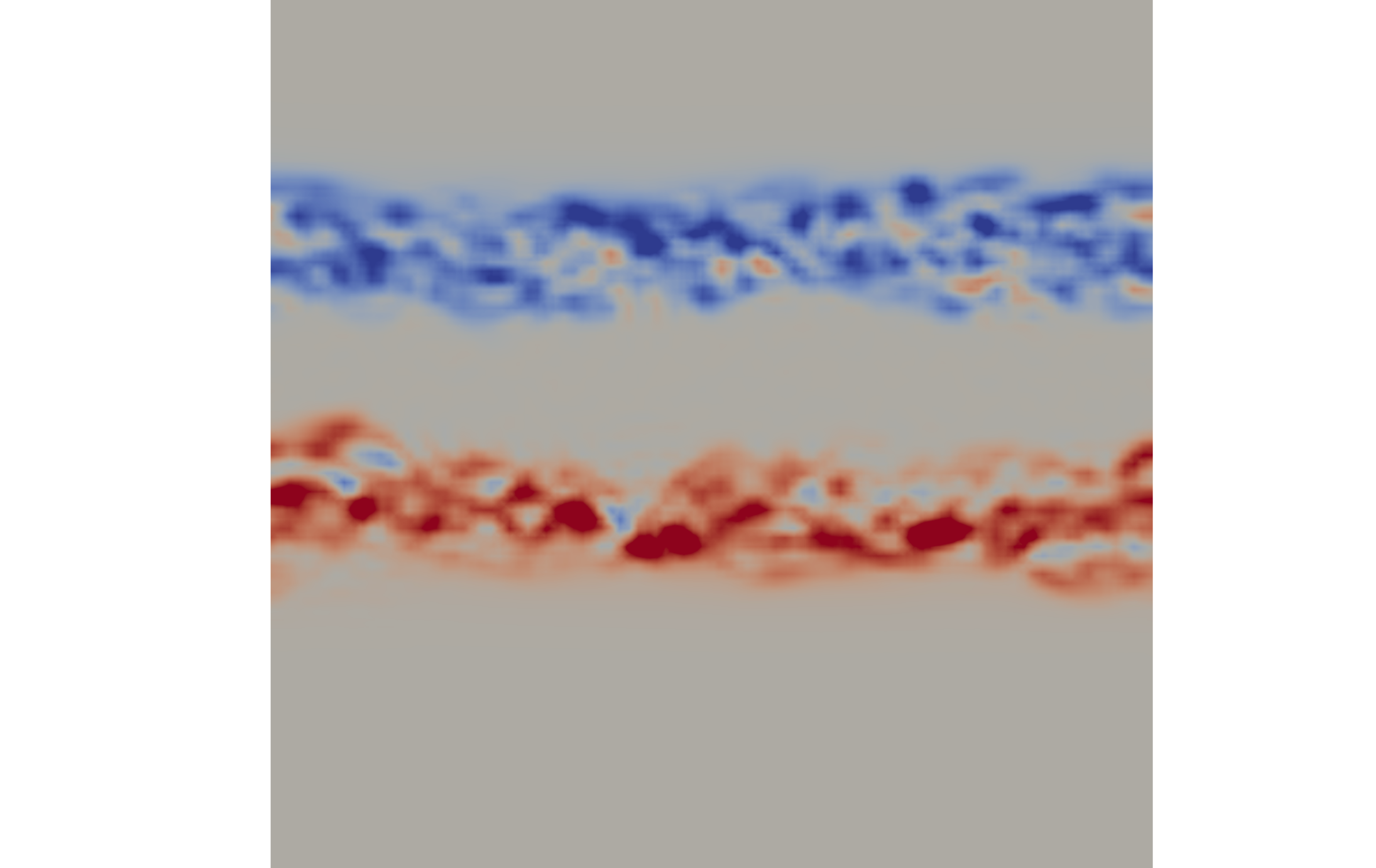}
        \put(30,60){}{}{(b)}
        \end{overpic}
    \end{subfigure}
    \hfill
    \begin{subfigure}[t]{0.30\textwidth} \centering
         \begin{overpic}[width=\textwidth]{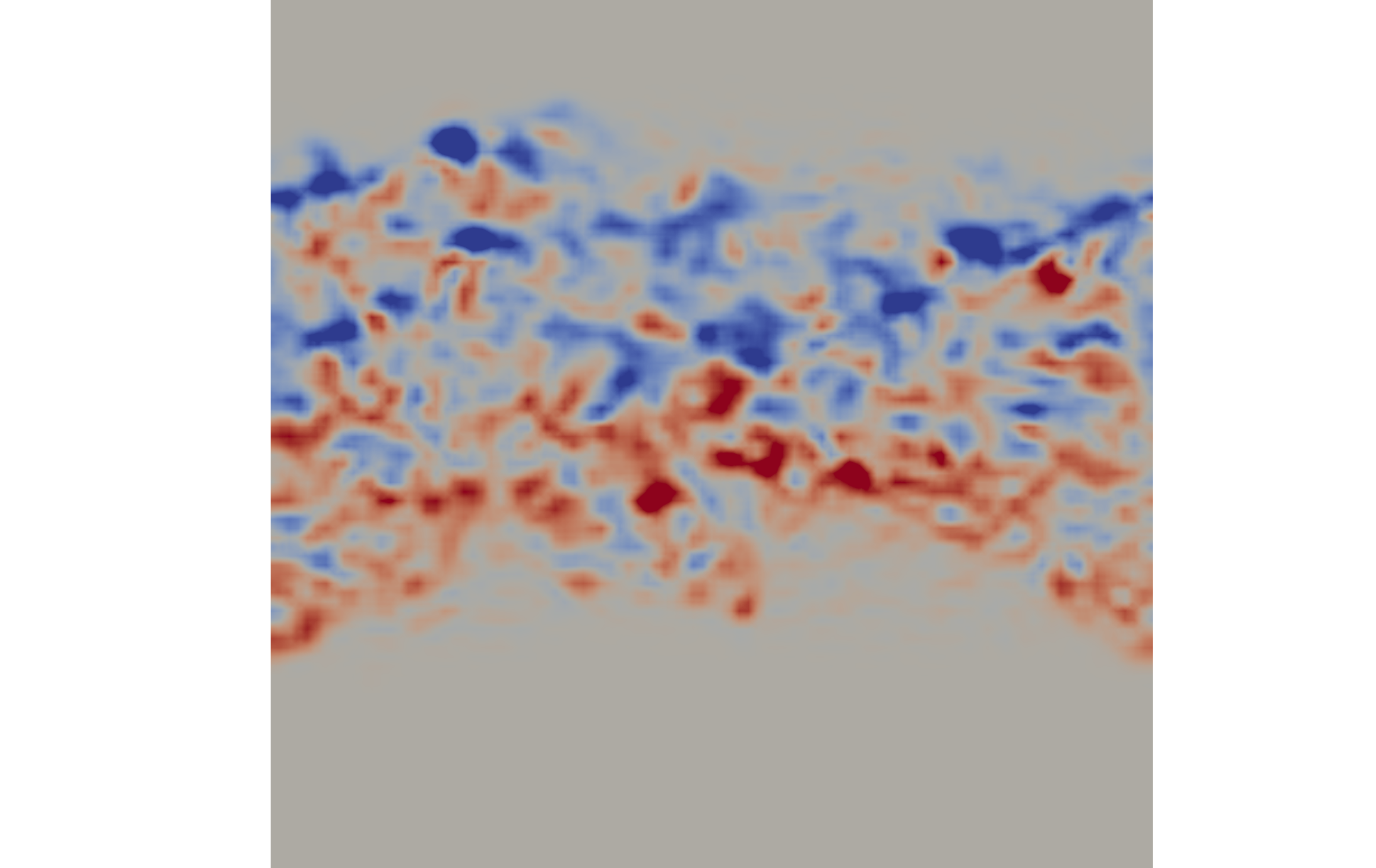}
        \put(30,60){}{}{(c)}
        \end{overpic}
    \end{subfigure}
\caption{Visualizations of transverse vorticity across the initially idealized jet as it becomes turbulent at \textit{(a)} $t_{\text{jet}}=$ 20, \textit{(b)} $t_{\text{jet}} =$ 30, and \textit{(c)} $t_{\text{jet}}=$ 42.}
\label{fig:vort_y}
\end{figure*}

The development of the jet is shown through transverse vorticity snapshots in \cref{fig:vort_y}.
We normalized the time as in Paoli et al. \cite{paoli2003dynamics}, where $t_{\text{jet}} = t \cdot V_{\text{jet}} / r_{\text{jet}}$,
and the resulting vorticity profiles are quite similar for the same non-dimensional times shown by the authors.
As soon has the jet has developed and lost more of its axial momentum, we use this solution as an initial condition for adding the vortex, assuming a fully rolled-up vortex, described  as the next simulation phase.

\subsection{Jet-vortex interaction phase} \label{sec:results_jet_vort_interaction}

In the present approach, we simulate the initial part of the vortex phase, in which the jet is entrained into the vortex as it loses its remainder axial momentum.
The subsequent part in which the vortex, i.e., the contrail plume, descends further and develops three-dimensional instabilities as the interaction with the atmosphere dynamics becomes more dominant, will be studied in future publications.

\begin{figure*}[hbt!] \centering
    \begin{subfigure}[t]{0.47\textwidth} \centering
        \begin{overpic}[width=\textwidth]{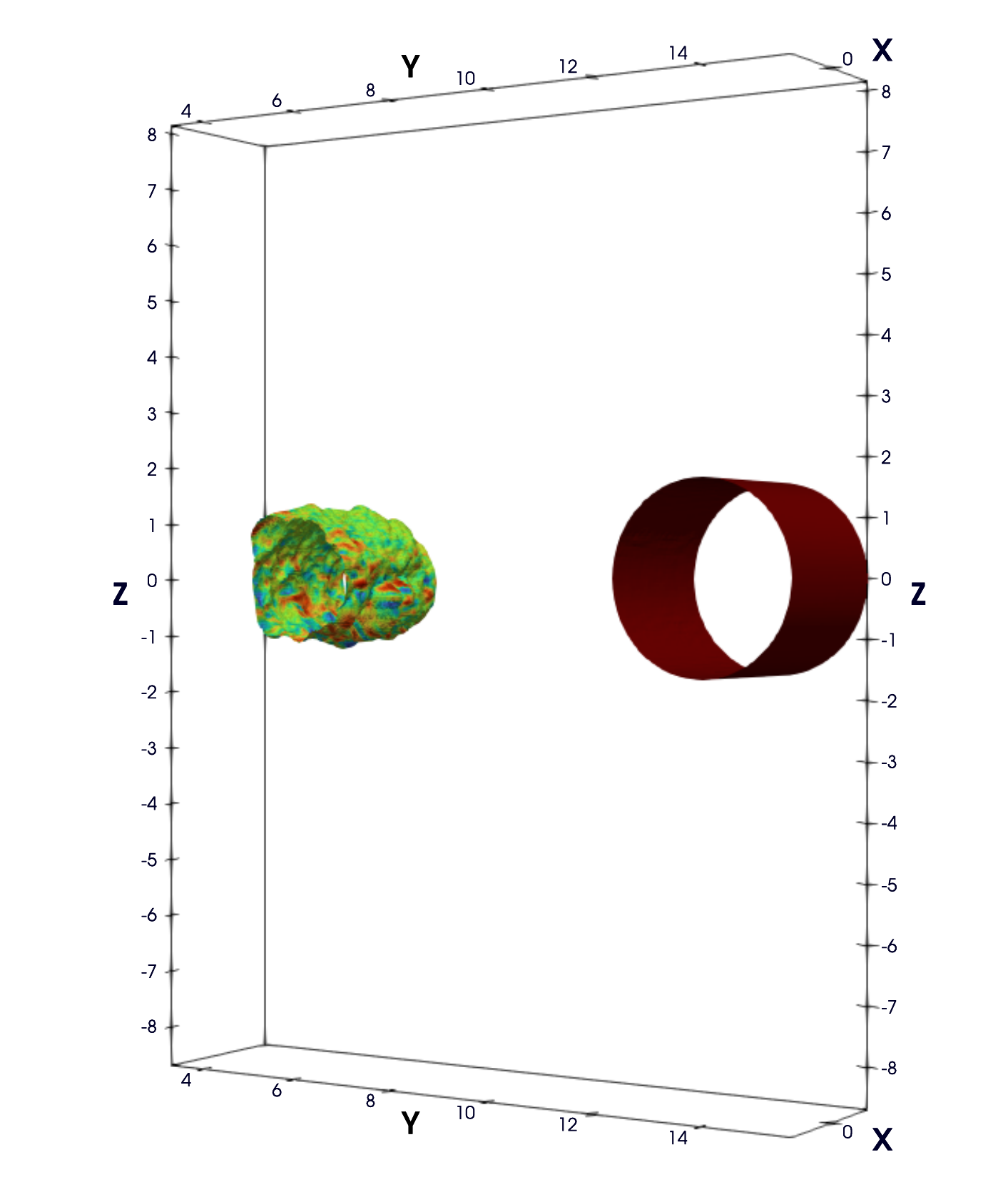}
            \put(30,60){}{}{(a)}
        \end{overpic}
    \end{subfigure}
\hfill
    \begin{subfigure}[t]{0.47\textwidth} \centering
        \begin{overpic}[width=\textwidth]{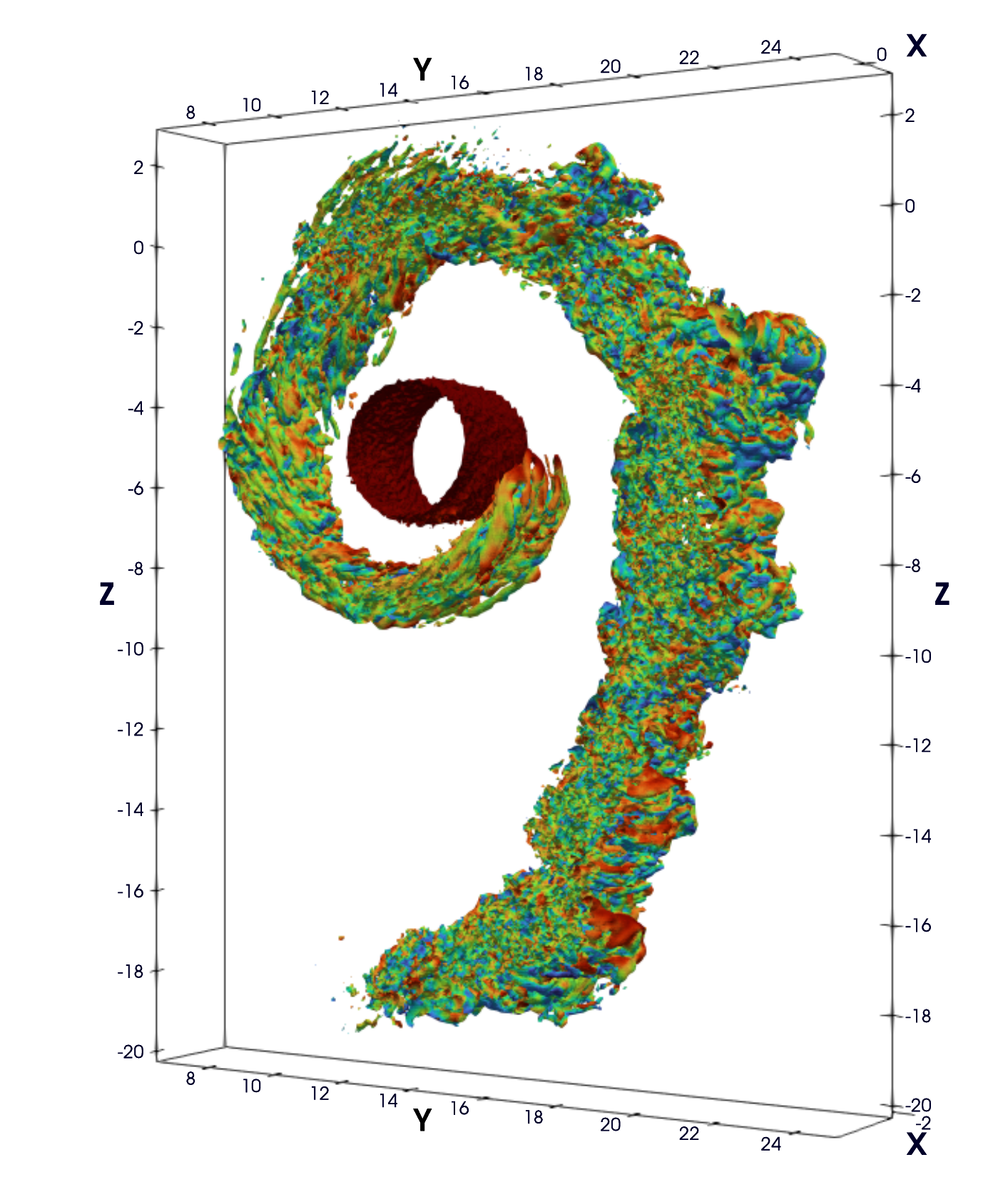}
            \put(30,60){}{}{(b)}
        \end{overpic}
    \end{subfigure}
 \caption{Visualization of jet and vortex interaction: iso-surface of vorticity magnitude colored by axial vorticity at \textit{(a)} 0.1 seconds and \textit{(b)} 5 seconds.} \label{fig:viz_vort}
    \end{figure*}

 \begin{figure*}[hbt!] \centering
    \begin{subfigure}[t]{0.47\textwidth} \centering
        \begin{overpic}[width=\textwidth]{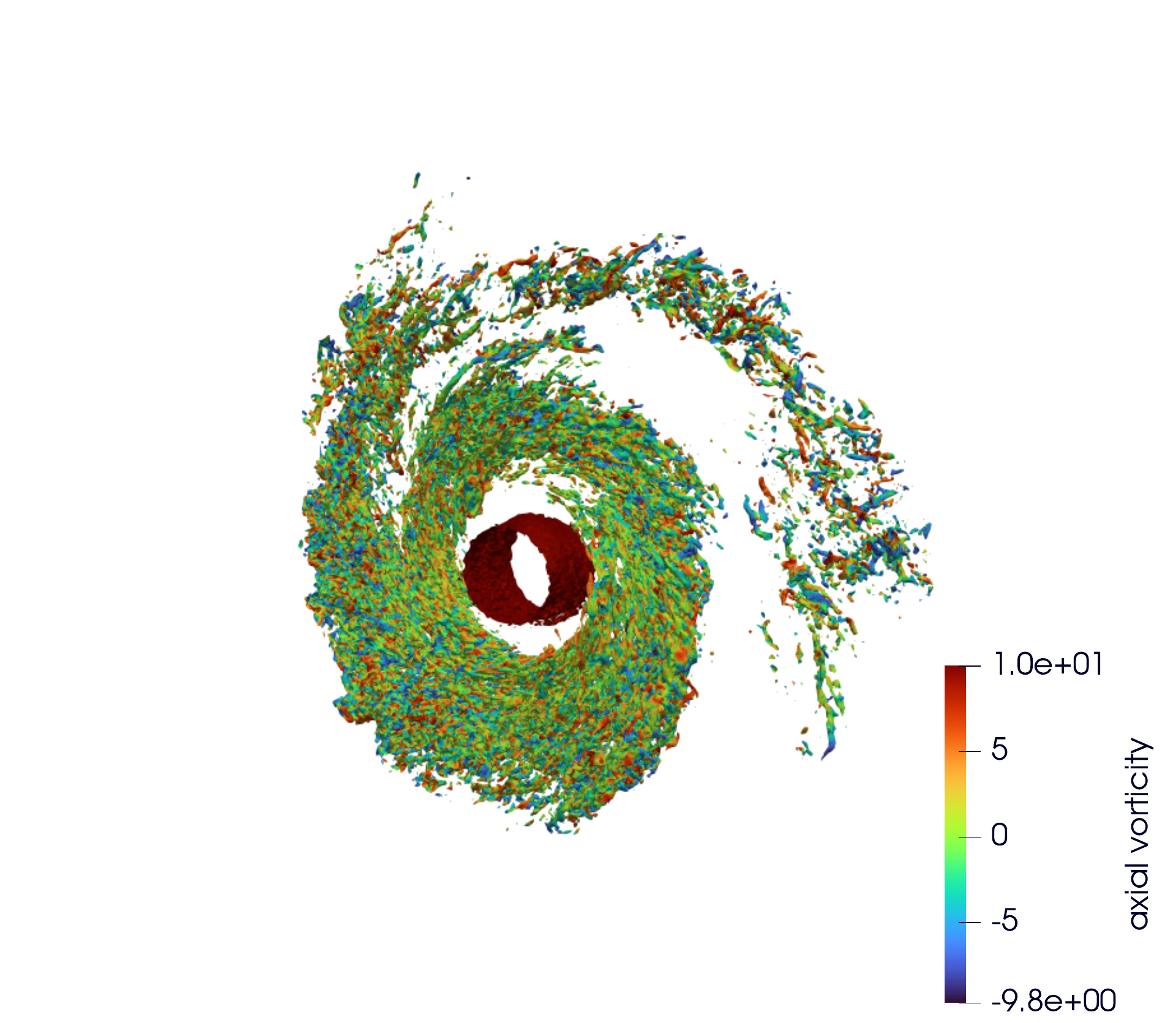}
            \put(30,60){}{}{(a)}
        \end{overpic}
    \end{subfigure}
\hfill
    \begin{subfigure}[t]{0.47\textwidth} \centering
        \begin{overpic}[width=\textwidth]{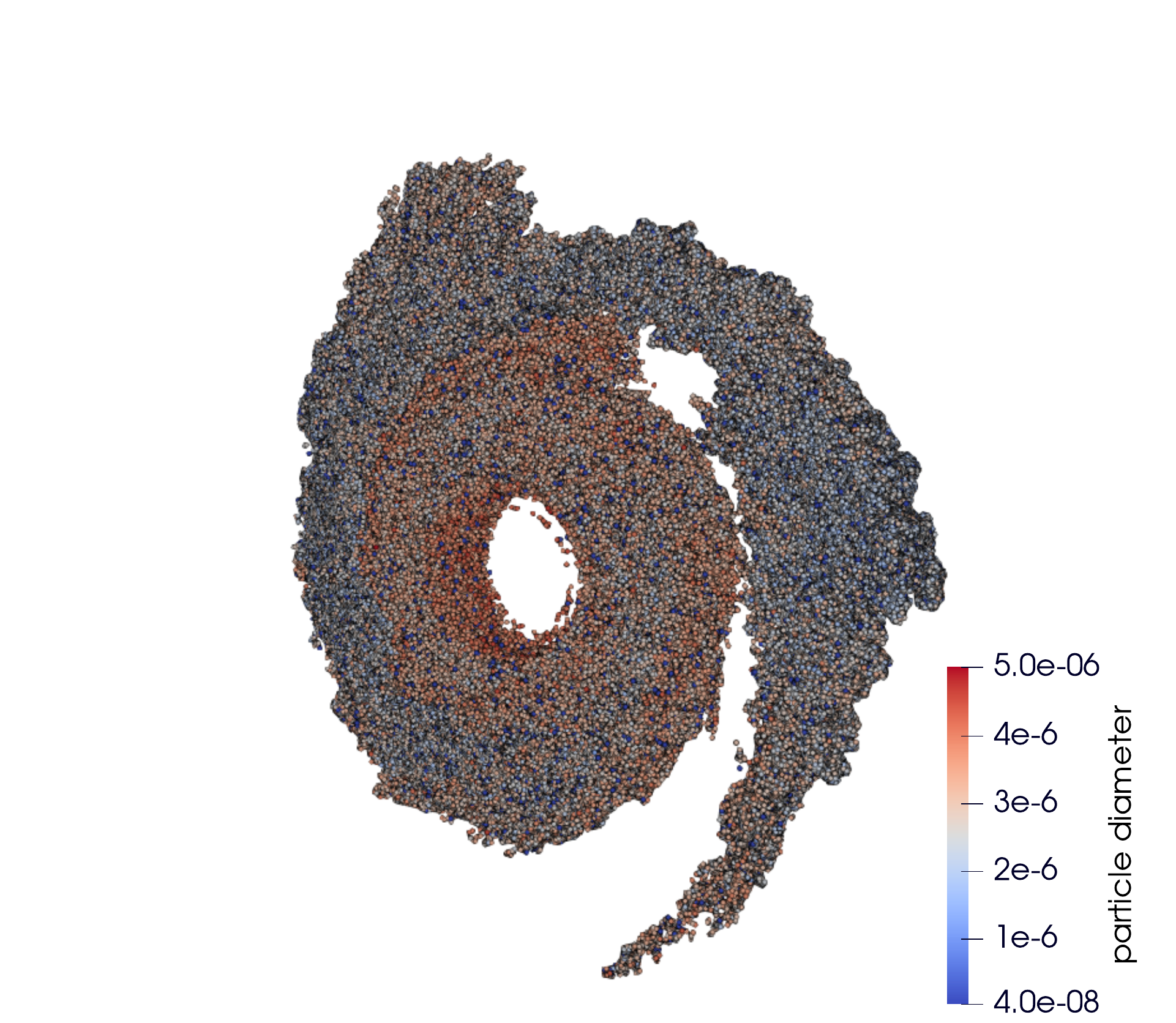}
            \put(30,60){}{}{(b)}
        \end{overpic}
    \end{subfigure}
 \caption{Visualization of vortex entrainment: \textit{(a)} iso-surface of vorticity magnitude colored by axial vorticity at 10 seconds and \textit{(b)} particle distribution around the vortex colored by particle diameter at 10 seconds.} \label{fig:viz_vort_dp}
\end{figure*}

\begin{figure*}[hbt!] \centering
    \begin{subfigure}[t]{0.47\textwidth} \centering
        \begin{overpic}[width=\textwidth]{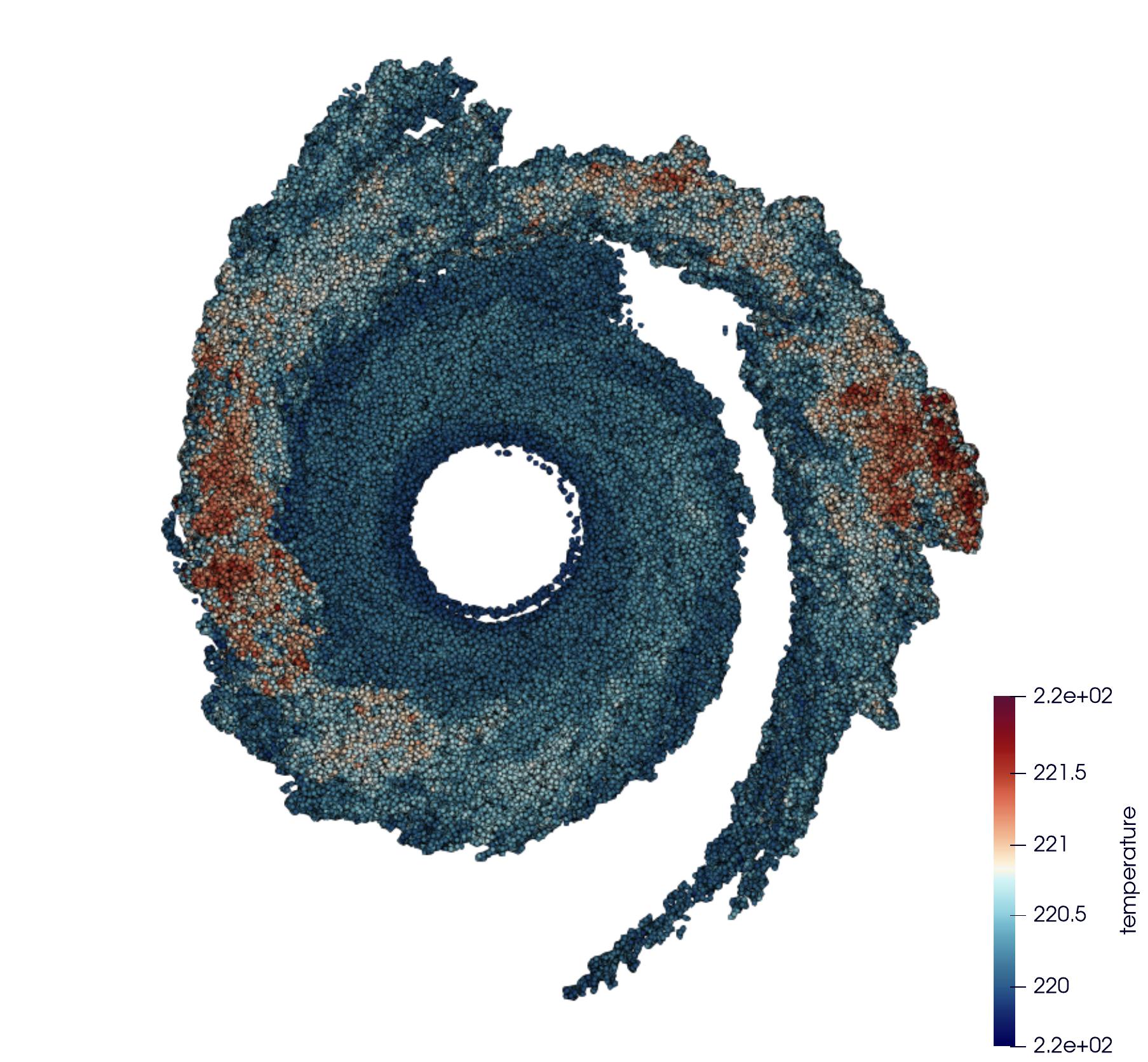} \put(30,60){}{}{(a)} \end{overpic}
    \end{subfigure}
\hfill
    \begin{subfigure}[t]{0.47\textwidth} \centering
        \begin{overpic}[width=\textwidth]{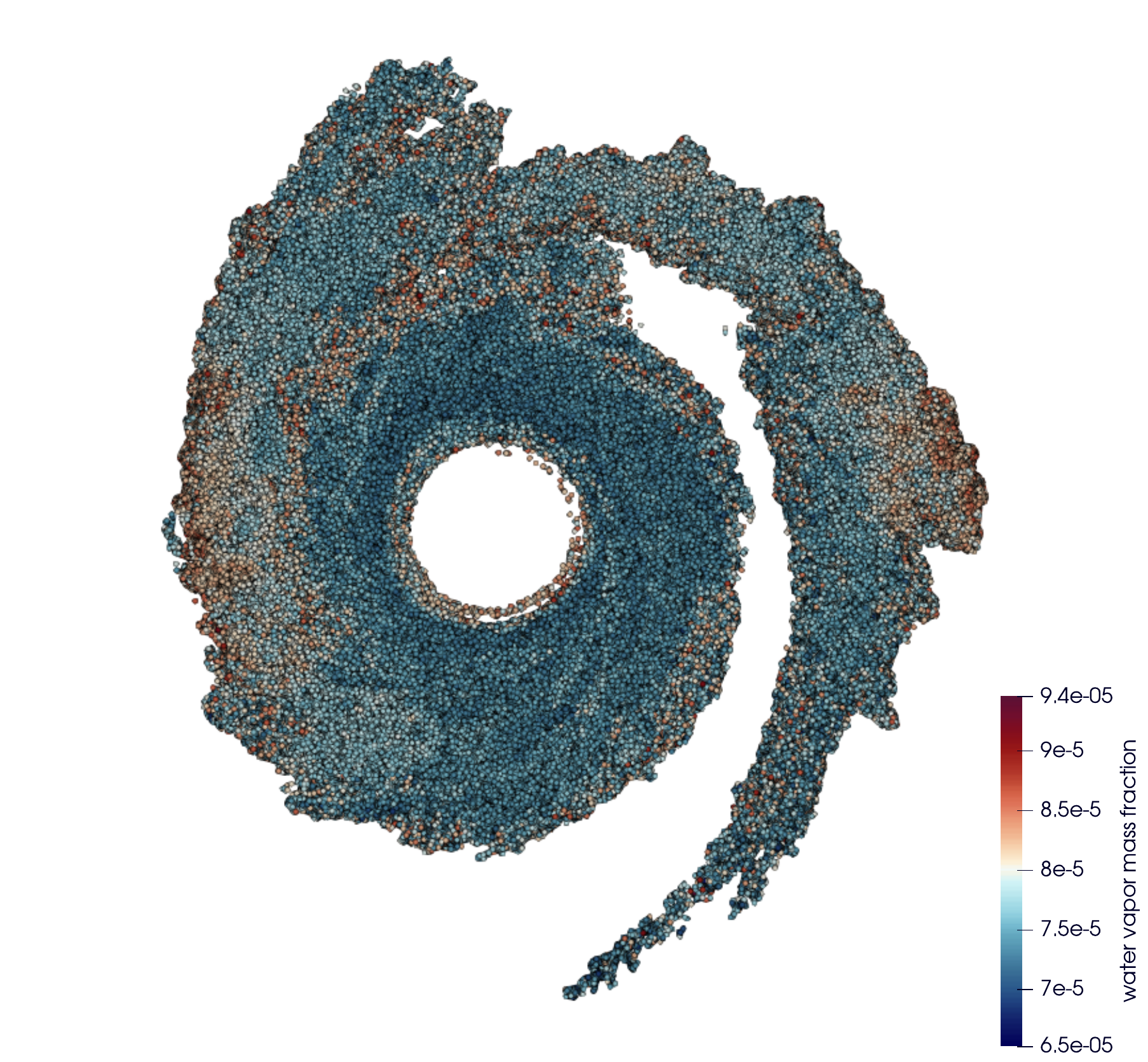} \put(30,60){}{}{(b)} \end{overpic}
    \end{subfigure}
\caption{Visualization of vortex entrainment: particle distribution around the vortex colored by surrounding \textit{(a)} flow temperature and \textit{(b)} water vapor mass fraction at 10 seconds.} \label{fig:lsp_prop_10sec}
\end{figure*}

In practice, a Lamb-Oseen vortex is introduced at a representative distance from the jet, as shown in \cref{fig:viz_vort} (a) -- as a result, the vortex will entrain the jet, which will also destabilize the vortex, as the iso-surface of vorticity magnitude shows in \cref{fig:viz_vort} (b) and  \cref{fig:viz_vort_dp} (a) . 
The entrainment is complete at around 10 seconds ($t_{\text{jet}} \approx $ 5000), 
As for the microphysics, as the water vapor laden jet cools down, saturation with respect to water is reached and the particles, now assumed as ice crystals, are allowed to grow according to a simple diffusion law described in \cref{sec:ice_model}.

We can observe in \cref{fig:viz_vort_dp} (b) a concentration of larger ice crystals close to the vortex core. This is because the temperature is lower at the core due to the pressure drop.
As a consequence, for the same quantity of water vapor, the supersaturation will be larger, and hence the ice crystals growth rate.
In fact, in \cref{fig:lsp_prop_10sec} (a), the temperature is lower around the core of the vortex, and there is also a concentration of water vapor around this region, cf. \cref{fig:lsp_prop_10sec} (b), which enables larger particle growth at the center of the plume.

The following sensitivity analysis follows the same core modeling and physical behavior displayed in these first two sections for the simulated jet and vortex interaction phases. The difference is that we simulate the fully domain with a jet and vortex pair and outlet and periodic boundary conditions for the lateral directions.

\subsection{Sensitivity to meshing, domain, and particles} \label{sec:sensitivity_meshing}

We analyze the impact of refining the mesh, extending the simulation domain in the axial direction, and increasing the number of computational particles on the jet decay, particle mean radius and radius distribution.
This study was made with the initial baseline, using ice microphysics and no bypass.
We summarize the outcomes in terms of the final relative difference in the main averaged variables in \cref{tab:meshing}.
The finer mesh has 20 points across the jet for the jet phase, and 32 points across the vortex for the jet-vortex interaction phase. As we did not observe significant changes with the mesh refinement, and further refining the mesh to 24 points per jet and 40 points per vortex did not enhance the difference, we use the baseline mesh as described \cref{sec:IC_BC} for the sensitivity analysis.

\begin{table}[hbt!]
\caption{\label{tab:meshing} Average sensitivity to meshing, domain, and particle number changes.}
\centering
\begin{tabular}{lcccc}
\hline
\textbf{}                  & \multicolumn{4}{c}{Relative difference with respect to baseline {[}\%{]}} \\ \cline{2-5}
Variable                   & Finer mesh        & Double $N_c$        & Seeding       & Double $L_x$ \\
\hline
Ice crystal number         & 0.1               & 0.0              & 0.5              & 0.2             \\
Mean ice crystal radius    & 1.3               & 0.0              & 0.5              & 0.9             \\
Relative humidity over ice & 0.4               & 0.0              & 0.1              & 0.1             \\
Pressure                   & 0.0               & 0.0              & 0.0              & 0.0             \\
Temperature                & 0.0               & 0.0              & 0.0              & 0.1             \\
\hline
\end{tabular}
\end{table}

Doubling the number of computational particles $N_c$ from approximately \num{0.5e6} to \num{1.0e6} results in a negligible difference, as we had previously seen when increasing to \num{2.0e6}. This indicates that, with our current refinement and for this large concentration of $N_c$ in a small core jet area, \num{0.5e6} computational particles are sufficient.

The setup of the jet, as described in  \cref{sec:IC_BC}, relies on initial random axial velocity perturbations to transition from a top hat profile to a turbulent Gaussian profile jet. 
Changing the cluster or the number of allocated nodes led to variability in the simulations.
We found the small difference in these perturbations to be due to a different seeding. This leads to variability in the jet development, as expected for turbulent flows.
In the average sense, we observed differences of 1.3\% in the mean particle radius by changing the number of allocated CPUs.
As such, we consider an averaged variability of this order to be within the variability of the jet, as verified in \cref{tab:meshing} for the above cases and also when doubling the axial length of the domain to $L_x=24r_j$.

. The tendency is for the finer mesh growing tail to lean towards large ice crystal size likely due to a small dissipation, but so if the case when the compute nodes are different, so we find this effect to be within the range of variability of the jet.

\subsection{Sensitivity to modeling} \label{sec:sensitivity_modeling}

\begin{figure*}[hbt!] \centering
    (a)\hspace{-1em}
        \begin{subfigure}[t]{0.47\textwidth} \centering
            \includegraphics[width=\textwidth]{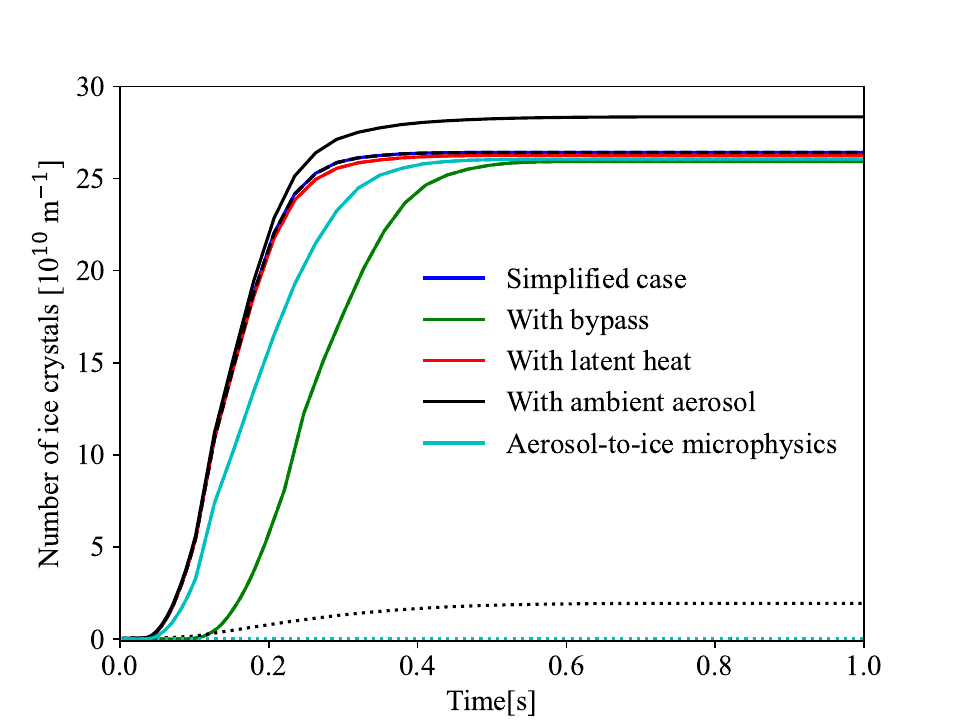}
        \end{subfigure}
    \hfill
    (b)\hspace{-1em}
        \begin{subfigure}[t]{0.47\textwidth} \centering
            \includegraphics[width=\textwidth]{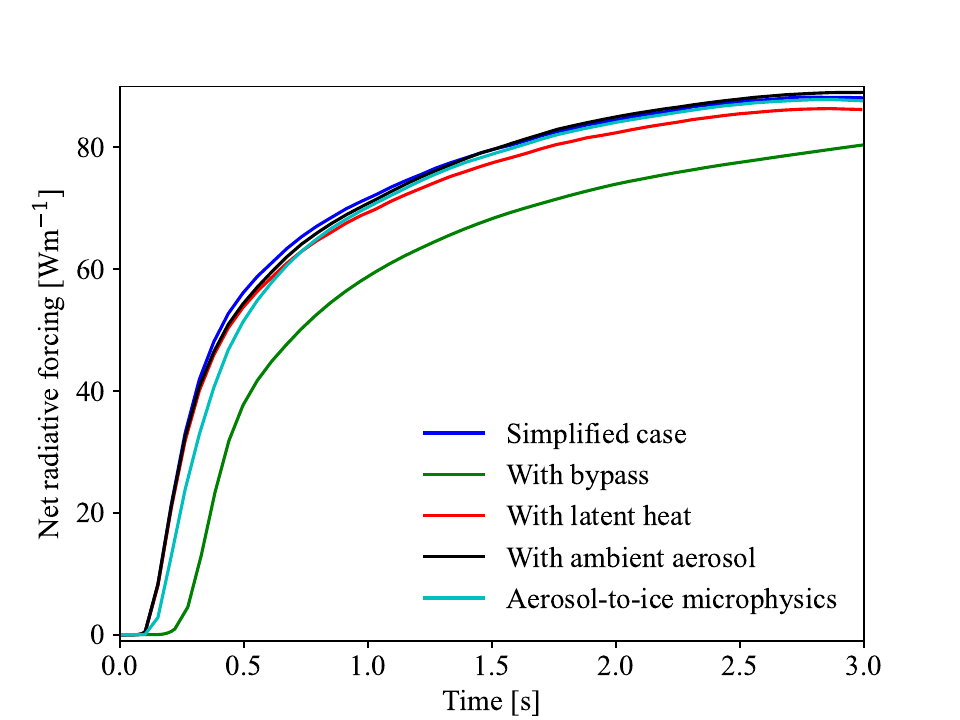}
        \end{subfigure}
\caption{Sensitivity to modeling choices: \textit{(a)} number of formed ice crystals, and \textit{(b)} net radiative forcing estimation per meter plume.}
\label{fig:sensitivity_modeling}
\end{figure*}

This early study had slightly different values from those used for the present baseline as described in \cref{sec:IC_BC}. As such, the sole purpose is to grasp the effect of modeling choices in temporal simulations of the near-field of contrails.
In this way, we add separately a representation of the bypass flow, latent heat, and ambient aerosol particles and extend the modeling from ice deposition only to aerosol-to-ice microphysics to our initially simplified case.
We note that the effect of each modeling change is relevant to the simplified case. 
The outcomes are presented in \cref{fig:sensitivity_modeling} in terms of the number of nucleated ice crystals and net radiative forcing. The net radiative forcing for all the cases presented here and henceforth is for a solar zenith angle of 0\si{\degree} and an albedo of 0.2, representative of noon over land.

\paragraph*{Adding latent heat}

Adding latent heat for condensational and depositional growth has a negligible effect on the activation of ice crystals, about 0.6\% less, and slightly reduces the ice crystal growth. This is due to the heat release effect on increasing the saturation pressure, leading to a slight difference in net radiative forcing of less 2\%.

\paragraph*{Adding co-axial jet -- bypass} \label{sec:results_bypass}

\begin{figure*}[hbt!] \centering
    (a)\hspace{-1em}
        \begin{subfigure}[t]{0.47\textwidth} \centering
            \includegraphics[width=\textwidth]{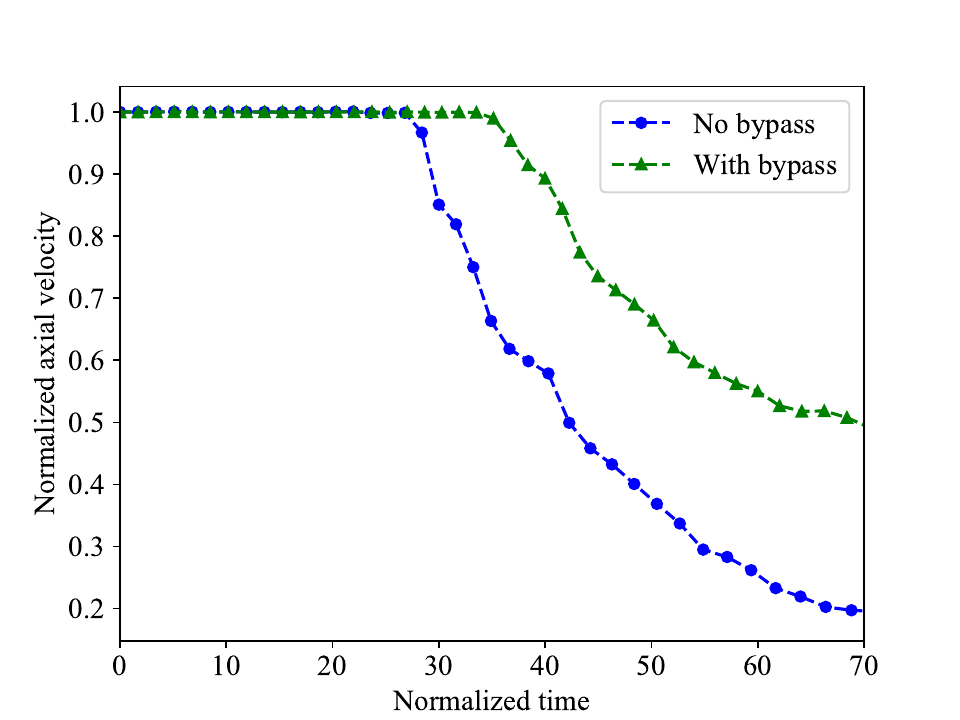} 
        \end{subfigure}
    \hfill
    (b)\hspace{-1em}
    \begin{subfigure}[t]{0.47\textwidth} \centering
        \includegraphics[width=\textwidth]{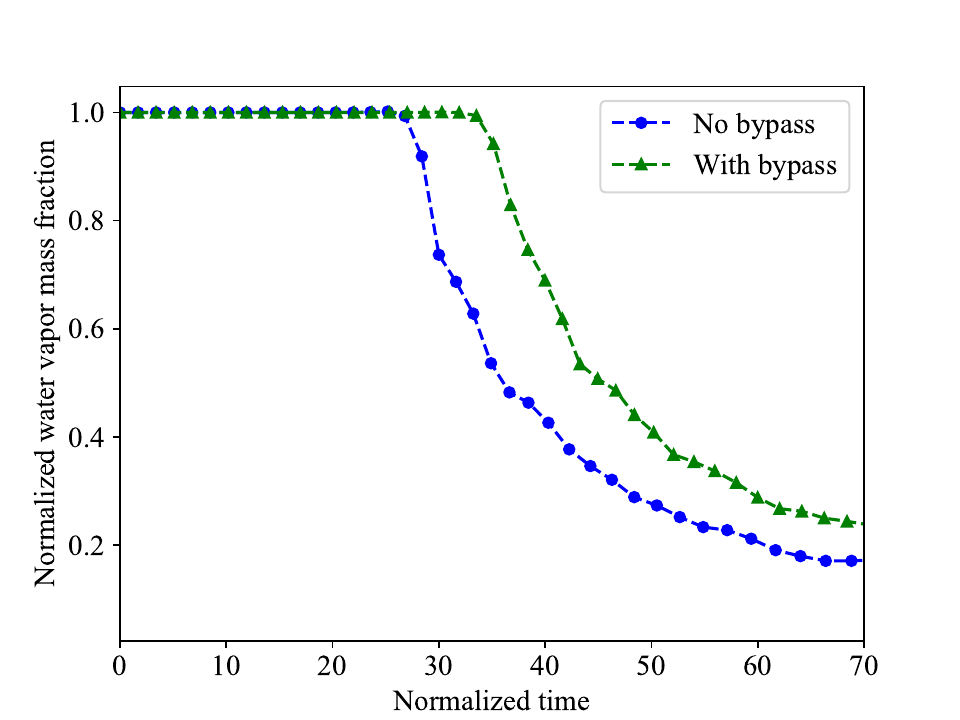}    
    \end{subfigure}
    \caption{Comparison of the jet decay between 
    cases \textit{(a)} without bypass (baseline), and \textit{(b)} with bypass.}
    \label{fig:jet_decay}
\end{figure*}

Adding a co-axial jet stabilizes the core hot flow and it takes longer to develop into turbulent and expand (a feature of the modeling). 
This is illustrated with the decay of the axial velocity and water vapor mass fraction averaged across the jet centerline in \cref{fig:jet_decay}. 
The decay takes longer to initiate for the case with the co-axial jet, i.e., bypass flow.
This delay also leads to a later relative humidity peak and this peak is slightly smaller due to the mixing that occurs in the mean time (not shown).
This is also true for the relative humidity over water, which controls the activation of particle growth, and the relative humidity over ice, which sets the rate of particle growth.

The evolution of the number of nucleated ice crystals in \cref{fig:sensitivity_modeling} (a) displays this delay for the bypass case and that less 2\% ice crystals form on soot particles. However, the decrease in net radiative forcing corresponds to a more significant value of 9\%, cf. \cref{fig:sensitivity_modeling} (b).
This can be explained thermodynamically: we are adding a band around the core flow with higher temperature air than ambient but the same quantity of water vapor. This results in lower relative humidity air, and thus, there is effectively less available vapor to deposit on the ice crystals.

\begin{figure*}[hbt!] \centering
    (a)\hspace{-1em}
        \begin{subfigure}[t]{0.47\textwidth} \centering
            \includegraphics[width=\textwidth]{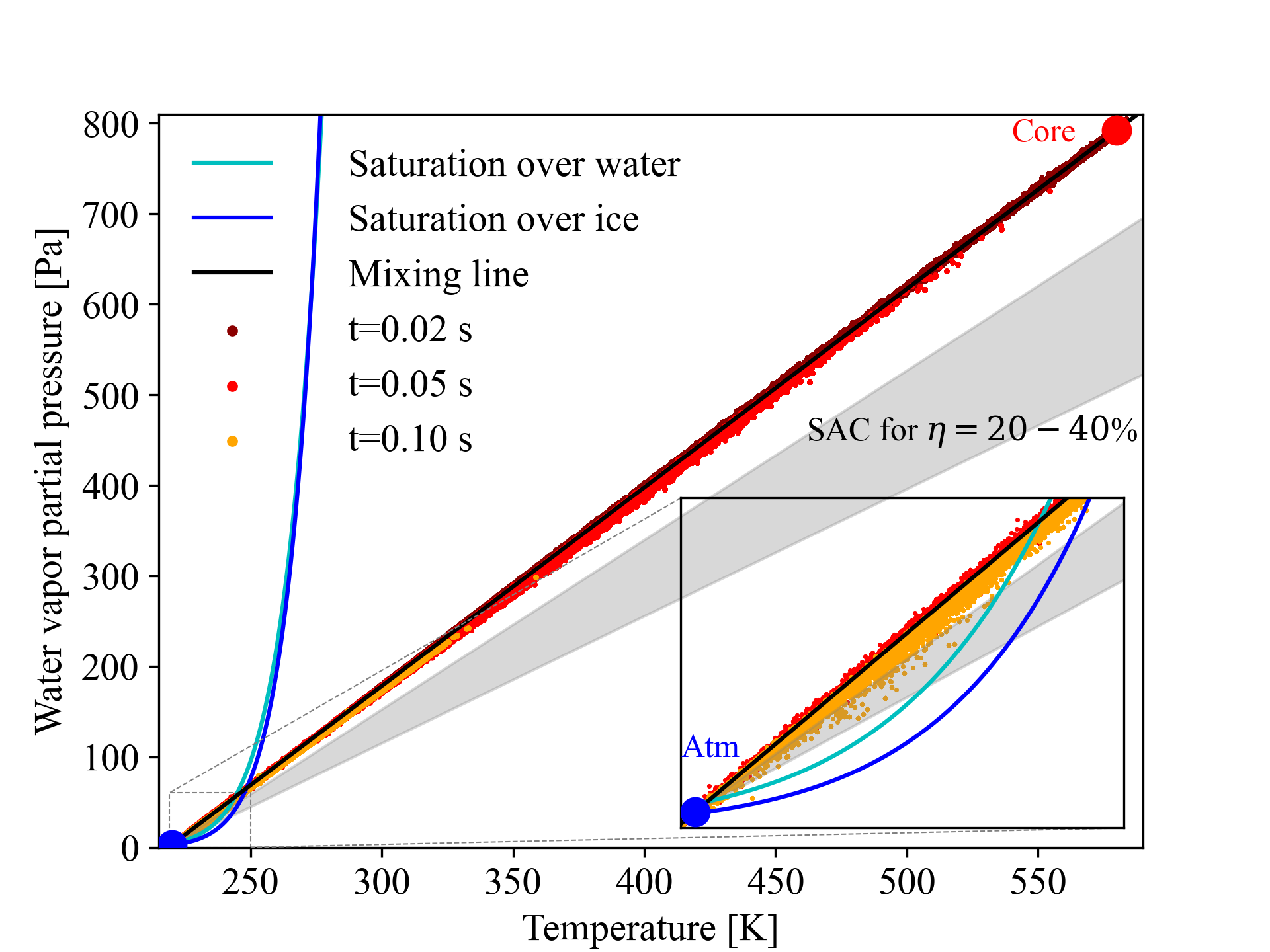} 
        \end{subfigure}
    \hfill
    (b)\hspace{-1em}
        \begin{subfigure}[t]{0.47\textwidth} \centering
             \includegraphics[width=\textwidth]{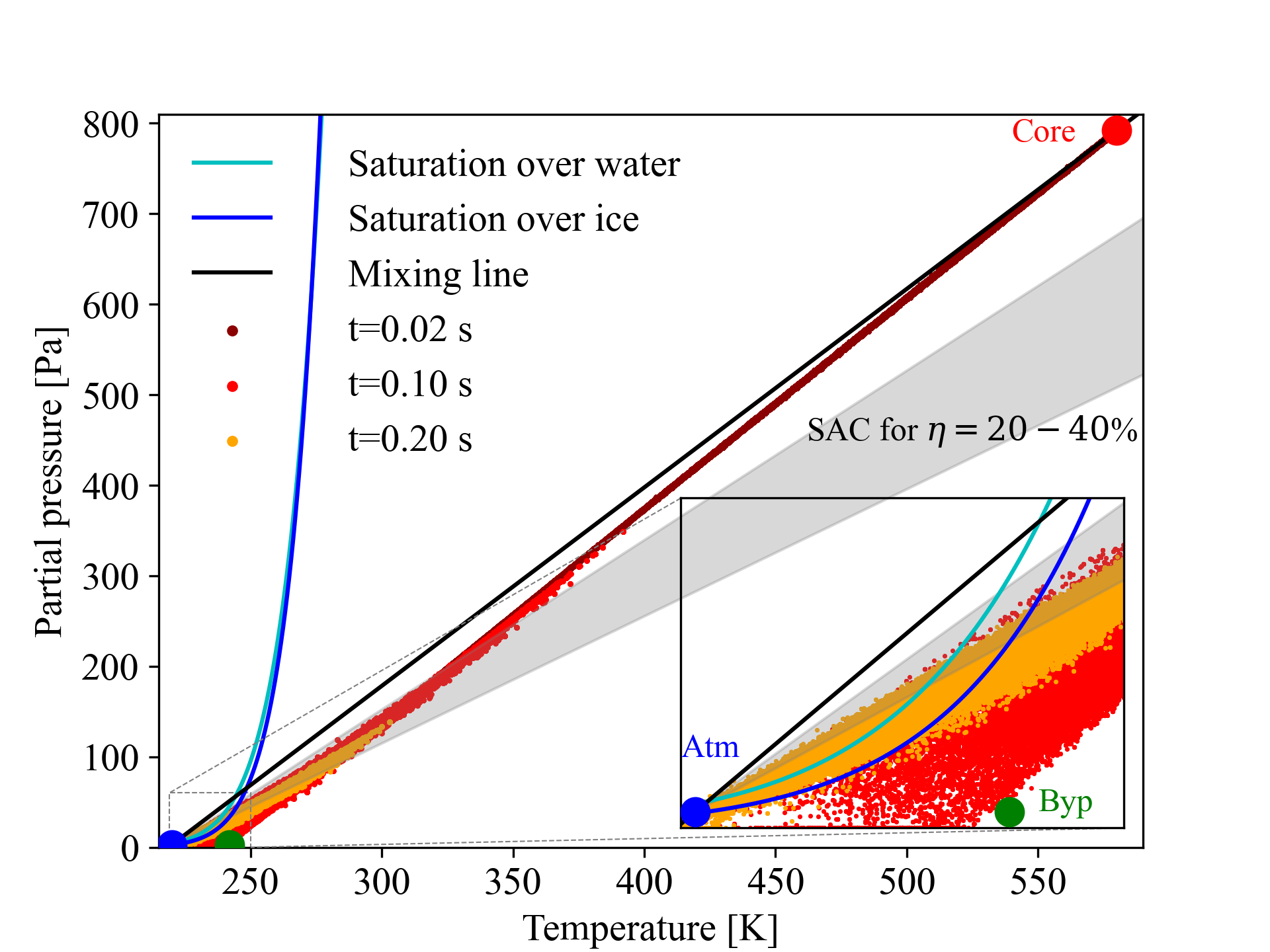} 
        \end{subfigure}
    \caption{Evolution of vapor pressure and temperature within the contrail for cases \textit{(a)} without bypass, and \textit{(b)} with bypass idealized flow.}    \label{fig:pv_bypass}
\end{figure*}

We can better illustrate this feature through the vapor pressure and temperature plots in \cref{fig:pv_bypass}.
The water vapor partial pressure versus temperature also shows a deviation from the mixing line based on the atmospheric and jet core values due to the less vapor-saturated bypass flow (assuming the same water vapor as the atmosphere but at a higher temperature).
This is in line with the study of Schumann et al. \cite{schumann1997large} where the differing heat and vapor emissions from the core and bypass flows lead to decorrelated temperature and vapor pressure values in the early contrail plume \cite{schumann1998dilution}.
As such, for threshold conditions for the formation of contrails, not all the particles may cross the water saturation curve and, hence, activate into water droplets.
This is due to the stronger entrainment of cold, less humid air that is warmer than atmospheric air, which leads to deviation from the mixing line at lower temperatures.
A more detailed study on this effect and the link to the Schmidt-Appleman criterion is nevertheless necessary.

In sum, modeling an idealized bypass profile leads to delayed jet development and a deviation from the heat and water vapor mixing line.
This may result in a smaller number of nucleated ice crystals, which will grow to larger sizes over time.

Though adding the bypass flow leads to a more realistic portrayal of an aircraft engine exhaust, this remains an idealized temporal approach which neglects the axial compression of the jet flow.
Cantin et al. \cite{cantin2022eulerian} have performed a spatial simulation with URANS of a detailed bypass and core jet flow and Khou et al. \cite{khou2015spatial,khou2017cfd} have used RANS to simulate the near-field of an aircraft including the core and bypass jet flows for early contrail formation.
Although a previous comparison between the temporal approach for both the jet near- and far-fields and spatial simulations of the jet flow yielded reasonable results in terms of parcel lifetime and dilution \cite{lewellen2020large}, it would still be interesting to compare present temporal simulations with geometry-bound spatial simulations of near-field contrails.

As a note, in our previous configuration \cite{ferreira2024developing}, we placed the particles within 50\% of the inner engine core radius -- this led to an enhanced difference in the rate of ice crystal nucleation (slope) when adding bypass as it delayed particle cooling. 
However, the influence of particle placement, whether within the full radius or half of it, on the rate of particle activation into ice crystals is minimal. 
In this past study, adding a bypass further reduces the number of nucleated ice crystals relative to the present case due to a lower water vapor mass fraction in the jet core of $Y_{\text{core}}=$\num{1.66e-2}.
This value had been computed following the SAC for an overall propulsive efficiency of 36\% and adjusting the size of the core radius. As such, the mixing line was closer to the saturation curve over water. Adding bypass flow led to a larger portion of particles not being able to reach the saturation criterion for activation into ice crystals due to the deviation from the mixing line.

\paragraph*{Varying the ice deposition coefficient}

This coefficient tunes the growth rate of the ice crystals -- a low coefficient (e.g., $\alpha=0.1$) leads to a slow growth in which most of the soot particles are able to activate into ice crystals uptake the available water vapor.
Using a larger coefficient (e.g., $\alpha=1.0$), the vapor depletion is so rapid that we have observed a reduction of almost 20\% in activated ice crystal numbers in past simulations based on the SAC for an engine efficiency around 36\% (not shown). 
Nevertheless, since with fewer particles there is less competition for water vapor, these grow to larger sizes. In this way, there is a catch up effect, and the resulting difference in net radiative forcing is minimal.
The effect is thus limited to near-field simulations of contrail formation.

\paragraph*{Adding ambient aerosol} \label{sec:results_lsp}

While previous cases mostly focus on soot emissions, we have added particles representing ambient aerosol with a concentration of 600 \si{\per\centi\meter}, monodisperse radius of 20 \si{\nano\meter} and hygroscopicity parameter $\kappa$ of 0.5 \cite{bier2024contrail}.
As a result, a portion of these ambient aerosol is able to activate into ice crystals and increase their total number by more than 7\%.
The number of ice crystals formed on ambient aerosol only becomes relevant after adding the vortex and entraining these particles in the plume.
Therefore, larger aircraft are expected to entrain more ambient aerosol and form even more ice crystals.
The effect on the net radiative forcing, cf. \cref{fig:sensitivity_modeling} (b) is not significant as the quantity of water vapor remains fixed.

\paragraph*{Adding aerosol-to-ice microphysics} \label{sec:results_micro}

The aerosol-to-ice microphysics approach includes activation of water droplets, condensation growth, homogeneous freezing, and deposition growth as described in \cref{sec:full_micro_model}.
We observe a limited effect on the final number of nucleated ice crystals in \cref{fig:sensitivity_modeling} (a) with respect to the baseline. The only striking difference is the slower rate of particle activation into ice crystals with respect to the ice deposition modeling only.
This is because, to become an ice crystal, the particle has to overcome the Kelvin effect, with a small help of the Raoult effect (due to the low assigned hygroscopicity for soot of $\kappa=$0.005), to become a droplet and experience a low enough freezing temperature.
In the ice microphysics modeling, the criterion is only to cross the saturation over water curve, so the particles become ice crystals more rapidly.
In the aerosol-to-ice, particles remain as water droplets in the first fractions of second.

Another interesting feature is that mean relative humidity over water and ice (not shown) depicts faster water vapor depletion rates for ice deposition modeling. 
This can also be explained because the ice crystal growth rate is larger than that of a water droplet at these temperatures, as the supersaturation over ice is higher than over water, yielding a more significant vapor gradient.
Nevertheless, we conclude that there is a limited effect in extending the microphysics modeling for these conditions, which leads to only a difference of minus 1\% in the net radiative forcing, cf. \cref{fig:sensitivity_modeling}.

We note that in the previous study \cite{ferreira2024developing}, the effect of adding aerosol-to-ice microphysics was assumed to impact the results more significantly. 
This was due to a bug in computing the cooling rate, leading to an artificially enhanced difference.

\subsection{Sensitivity to atmospheric, aircraft and engine parameters} \label{sec:sensitivity_ac_engine}

The baseline for this study has been described in \cref{sec:IC_BC} and includes the idealized bypass flow, aerosol-to-ice microphysics, latent heat, and ambient aerosol with a concentration of 600 \si{\per\cubic\centi\meter}.
The hygroscopicity parameter $\kappa$ is equal to 0.005 for soot particles and 0.5 for ambient aerosol, similarly to Bier et al. \cite{bier2024contrail}.

\paragraph*{Atmospheric parameters} \label{sec:sensitivity_atm}

\begin{figure*}[hbt!] \centering
    (a)\hspace{-1em}%
        \begin{subfigure}[t]{0.47\textwidth} \centering
            \includegraphics[width=\textwidth]{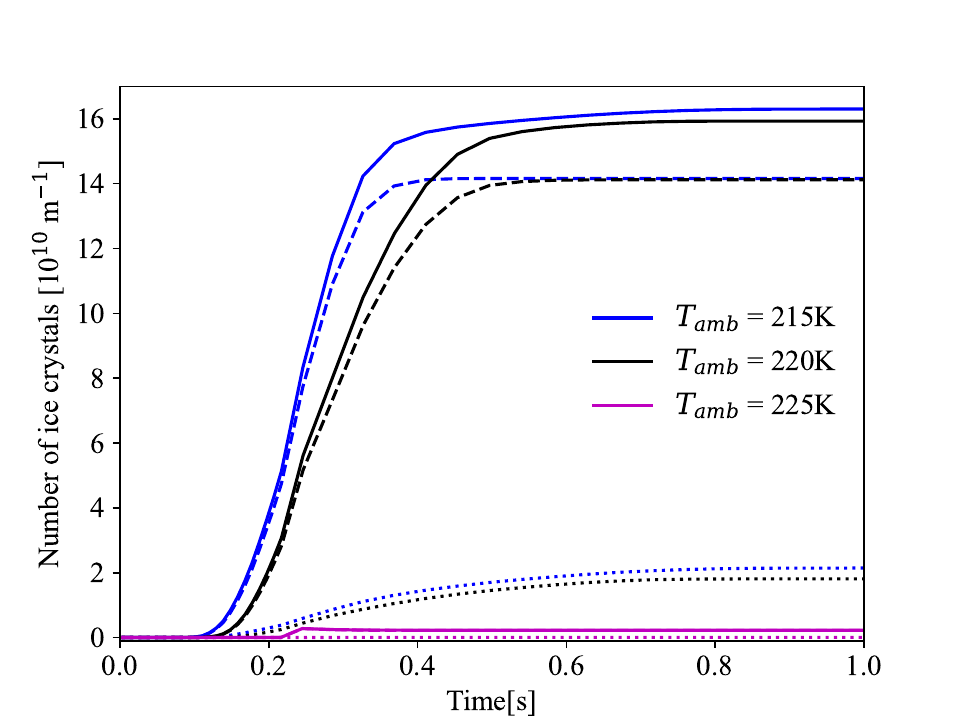} \label{fig:t_Tamb}
        \end{subfigure}
    \hfill
    (c)\hspace{-1em}
        \begin{subfigure}[t]{0.47\textwidth} \centering
            \includegraphics[width=\textwidth]{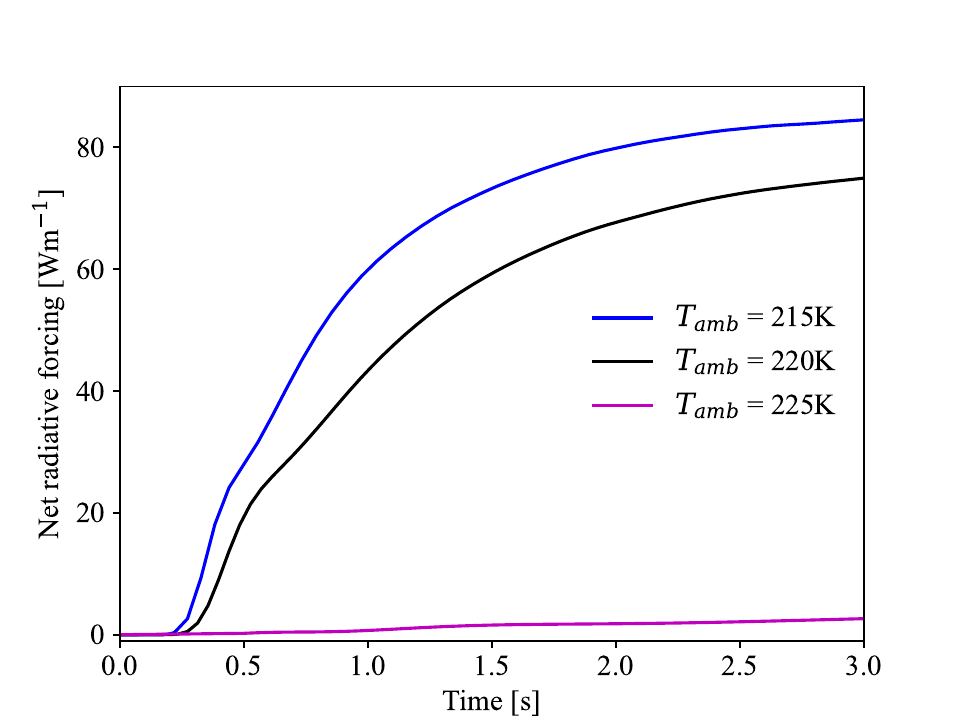}   \label{fig:rf_Tamb}
        \end{subfigure}
    \vskip\baselineskip
    (b)\hspace{-1em}
        \begin{subfigure}[t]{0.47\textwidth} \centering
            \includegraphics[width=\textwidth]{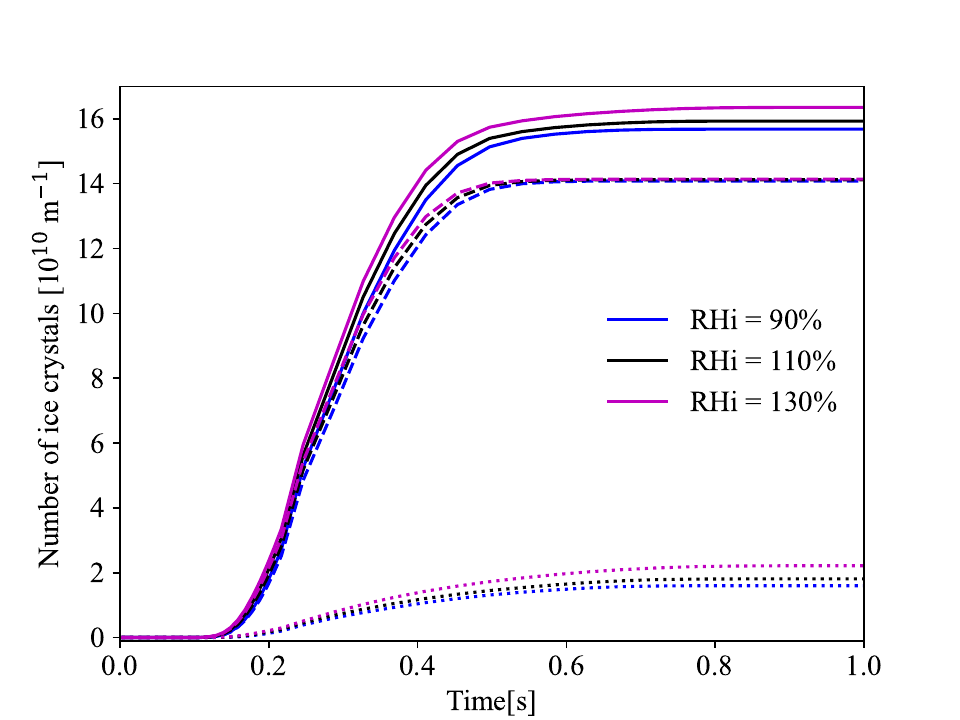}   \label{fig:t_rhi}
        \end{subfigure}
    \hfill
    (d)\hspace{-1em}
        \begin{subfigure}[t]{0.47\textwidth} \centering
            \includegraphics[width=\textwidth]{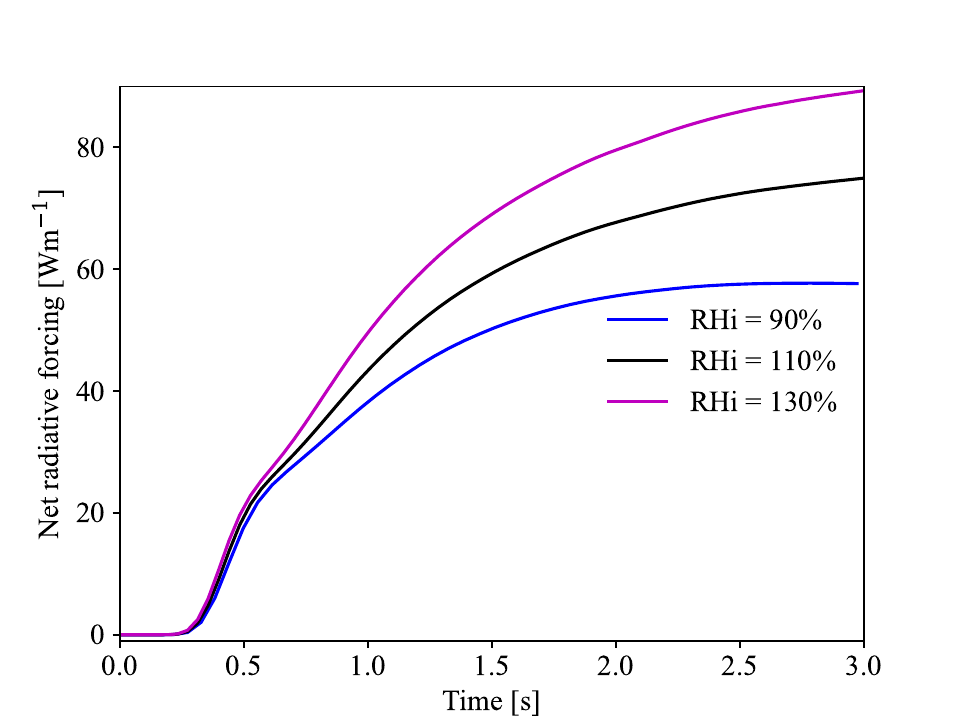} \label{fig:rf_rhi}
        \end{subfigure}
    \caption{Sensitivity to temperature, and relative humidity over ice for: \textit{(a-c)} total number of ice crystals (\rule[0.5ex]{.4cm}{0.4pt}), formed on soot particles (\rule[0.5ex]{.1cm}{0.4pt} \rule[0.5ex]{.1cm}{0.4pt} \rule[0.5ex]{.1cm}{0.4pt}) and on ambient aerosol ($\cdot\cdot\cdot$); \textit{(d-f)} and net radiative forcing per meter of plume.} \label{fig:results_atm}
\end{figure*}

We varied the atmospheric temperature from 215 \si{\kelvin} to 225 \si{\kelvin}, scaling the water vapor mass fraction for the same relative humidity over ice of 110\% and pressure of 240 \si{\hecto\pascal}.
Fewer ice crystals are formed on soot particles and ambient aerosol as the atmospheric temperature increases, as shown in \cref{fig:results_atm} (a). 
For the highest temperature, in which less than 5\% ice crystals form with respect to the baseline, the contrail is not visible as the 90\textsuperscript{th} percentile of the optical depth is below 0.02. (not shown).
The lowest temperature has 13\% higher net radiative forcing then the baseline, as shown in \cref{fig:results_atm} (c).

The variation of the atmospheric relative humidity over ice from 90\% to 130\% only affects the rate of the nucleated of ice crystals formed on soot but then the final number as saturation over water is easily reached within the plume, cf. \cref{fig:results_atm} (b). It does change the number of ice crystals formed on ambient aerosol as they are entrained with the (un-)saturated atmospheric air by the vortex.
The relative humidity has a lasting effect on the net radiative forcing of the contrail, cf. \cref{fig:results_atm} (d), as the growth of the ice crystals depends on saturation over ice. The difference with respect to the baseline is $-23/+19$\% at 3 \si{\second}, and it will keep on growing.

\paragraph*{Aircraft and engine parameters}

\begin{figure*}[hbt!] \centering  
    (a)\hspace{-1em}%
        \begin{subfigure}[t]{0.47\textwidth} \centering
            \includegraphics[width=\textwidth]{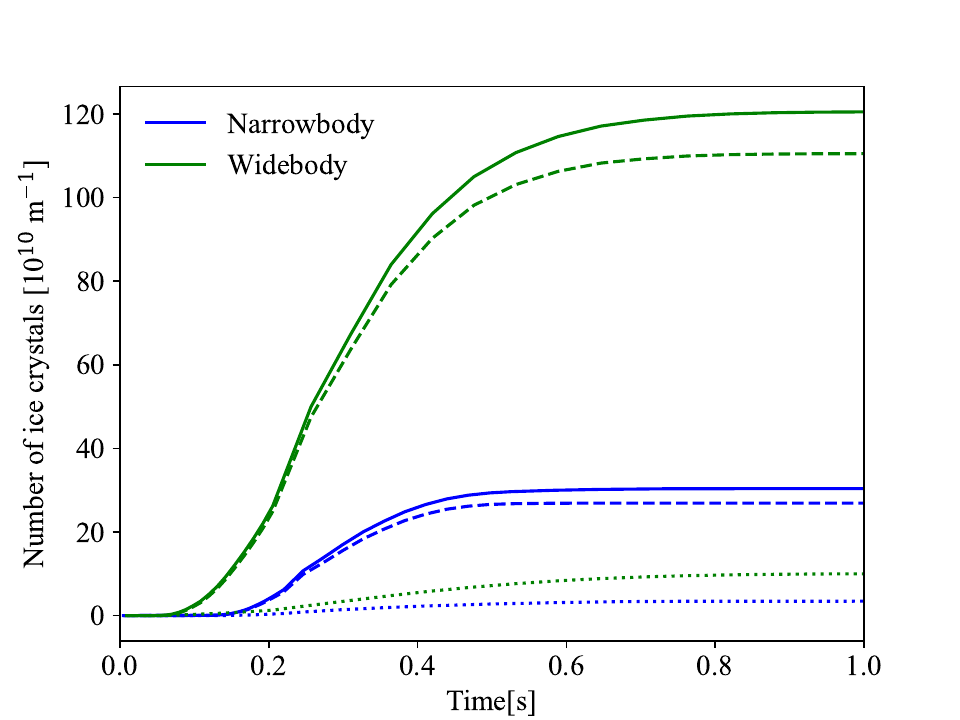}   \label{fig:t_AC}
        \end{subfigure}
    \hfill
    (d)\hspace{-1em}%
        \begin{subfigure}[t]{0.47\textwidth} \centering
            \includegraphics[width=\textwidth]{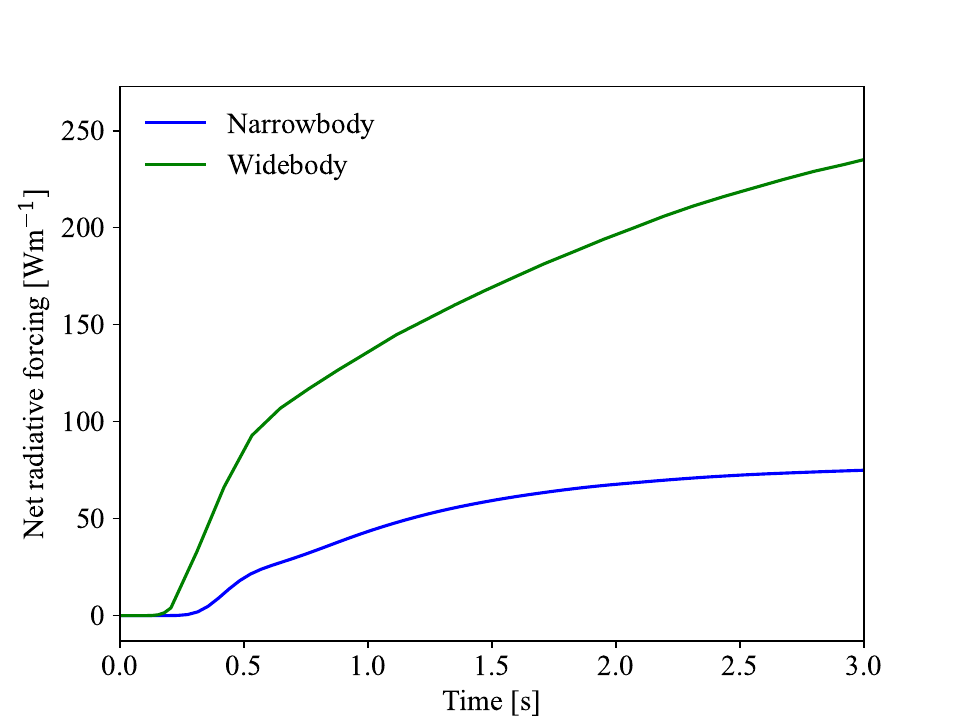}  \label{fig:rf_AC}
        \end{subfigure}
    \vskip\baselineskip
    (b)\hspace{-1em}%
        \begin{subfigure}[t]{0.47\textwidth} \centering
            \includegraphics[width=\textwidth]{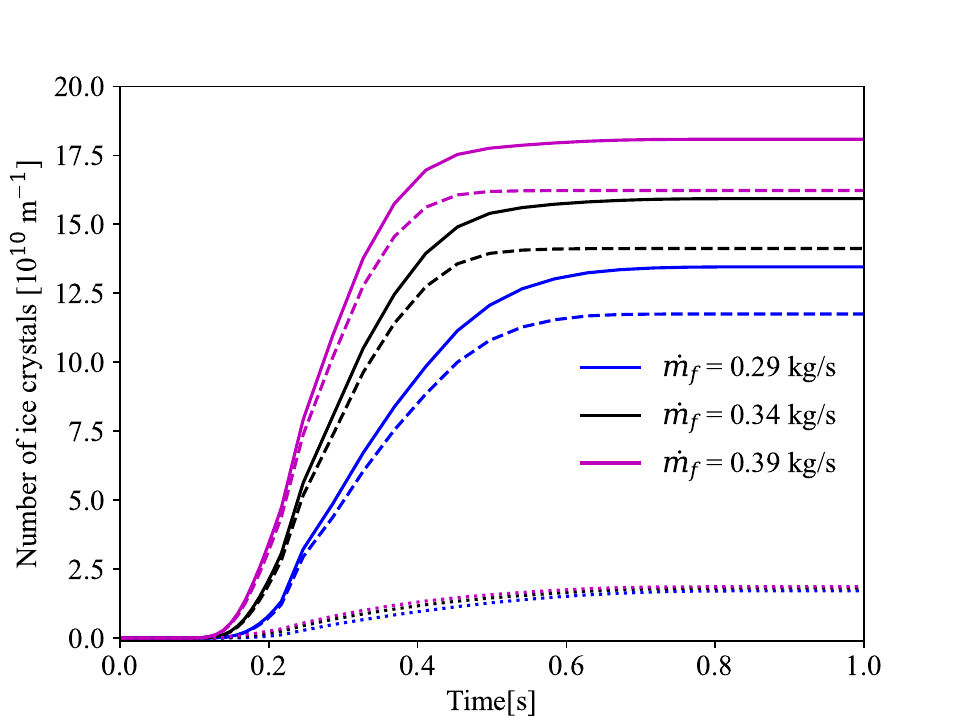} \label{fig:t_Y_jet}
        \end{subfigure}
    \hfill
    (e) \hspace{-1em}%
        \begin{subfigure}[t]{0.47\textwidth} \centering
            \includegraphics[width=\textwidth]{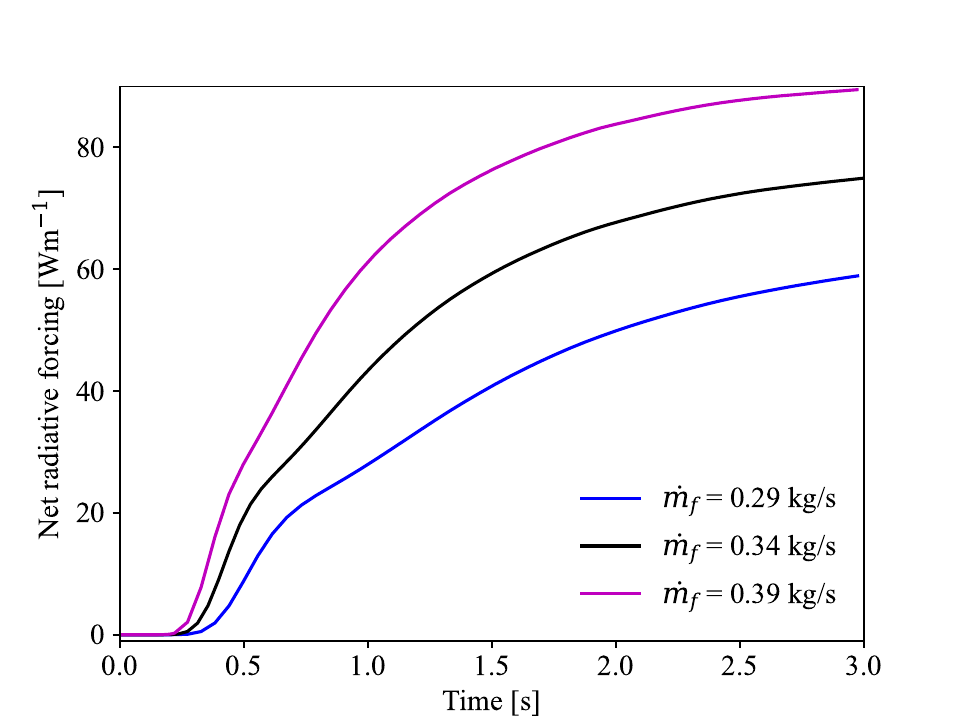}   \label{fig:rf_Y_jet}
        \end{subfigure}
    \vskip\baselineskip
    (c)\hspace{-1em}%
        \begin{subfigure}[t]{0.47\textwidth} \centering
            \includegraphics[width=\textwidth]{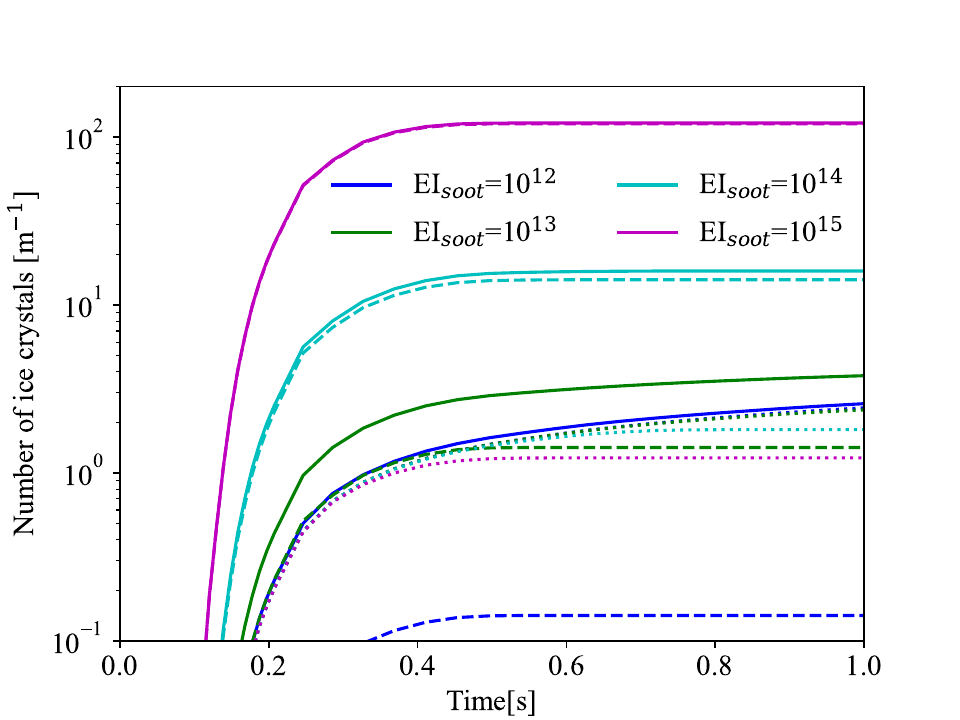}   \label{fig:t_soot}
        \end{subfigure}
    \hfill
    (f)\hspace{-1em}%
        \begin{subfigure}[t]{0.47\textwidth} \centering
            \includegraphics[width=\textwidth]{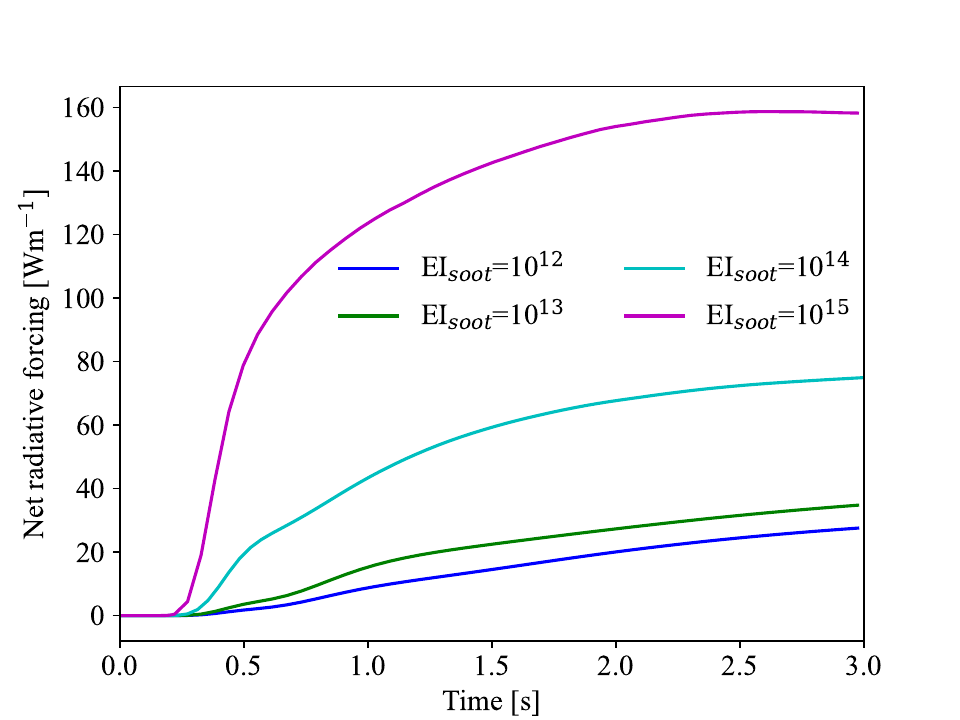}   \label{fig:rf_soot}
        \end{subfigure}
    \caption{Sensitivity to aircraft size, fuel rate, and soot number emission index for: \textit{(a-c)} total number of ice crystals (\rule[0.5ex]{.4cm}{0.4pt}), formed on soot particles (\rule[0.5ex]{.1cm}{0.4pt} \rule[0.5ex]{.1cm}{0.4pt} \rule[0.5ex]{.1cm}{0.4pt}) and on ambient aerosol ($\cdot\cdot\cdot$); \textit{(d-f)} and net radiative forcing per meter of plume.}  \label{fig:sensitivity_engine_AC}
\end{figure*}

In this section, we discuss the sensitivity to the size of the aircraft, fuel consumption, and soot number emission index. The results in terms of number of nucleated ice crystals and net radiative forcing per meter of plume are presented in \cref{fig:sensitivity_engine_AC}.
First, we analyzed the difference between a narrowbody (based on a B373/A320, \textit{baseline}) and a widebody (based on a B777/A350) representative aircraft. 
For the widebody, we assume a jet core radius $r_{\text{core}}$ of \num{0.61} \si{\meter} and a fuel consumption of 1.5 \si{\kilo\gram\per\second} at cruise.
The remaining parameters are scaled according to the narrowbody's same flight parameters and bypass ratio of about 8.
While a narrowbody's aircraft and engine performance differs from a widebody's, we aim to minimize the number of variables in this comparison.
Similarly to the narrowbody in \cref{tab:aircraft_narrowbody}, the aircraft and engine parameters for the widebody are summarized in \cref{tab:aircraft_widebody}. 

\begin{table}[hbt!] \centering
\caption{Properties of a representative widebody aircraft and engine.}
\label{tab:aircraft_widebody} 
\begin{tabular}{llll|llll}  \hline
\multicolumn{4}{c|}{Widebody aircraft}                       & \multicolumn{4}{c}{Engine}               \\  \hline
Flight velocity     & $V_{\text{a/c}}$  & 252  & \si{\meter\per\second}  & Fuel mass flow    &   $\dot{m}_{\text{f}}$                 & 1.50  & \si{\kilogram\per\second} \\
Circulation         & $\Gamma$          & 570    & \si{\meter\squared\per\second} & Soot emission index &  $\text{EI}_{\text{soot}}$   &  \num{e14}  & \si{\per\kilogram} fuel \\
Wingspan             & $b$              & 65 & \si{\meter}              & Water  emission index    &   $\text{EI}_{\text{H$_2$0}}$  & 1.25   & \si{\kilogram\per\kilogram} fuel \\
Vortex core distance & $b_0$            & 51   & \si{\meter}              & Water mass emission     &   $m_{\text{Y}}$               & 3.9 & \si{\gram\per\meter} \\
Vortex core radius   & $r_v$            &  4   & \si{\meter}              & Jet core radius   &  $r_{\text{core}}$  &    0.61   &\si{\meter}         \\
Jet/vortex distance  & $d_{\text{j/v}}$ & 15.5   & \si{\meter}     & Jet total radius  &  $r_{\text{total}}$   &   0.85  & \si{\meter}         \\\hline
\end{tabular}
\end{table}

As expected, the widebody forms more ice crystals as it emits 4.5 times more soot particles than the narrowbody aircraft (due to the larger fuel consumption).
In addition, given the bigger wingspan and mass of the aircraft, the vortical motion is stronger and entrains more ambient aerosol than the narrowbody. As a result, more than 2.9 times more ambient aerosol nucleate into ice crystals per meter, and 4.8 times more soot particles becomes ice crystals \cref{fig:sensitivity_engine_AC} (a). 
The presence of a higher number of hygroscopic aerosol competes with the lower hygroscopicity soot particles. Hence, we observe that a smaller fraction of ice crystals form on soot for the widebody with respect to the narrowbody baseline.
Finally, the effect on the net radiative forcing is 3 times higher for the widebody aircraft at 3 \si{\second}.

The effect of varying the fuel consumption rate is simulated by adjusting the water vapor mass fraction in the jet core and the number of soot particles accordingly.
As such, we can observe in \cref{fig:sensitivity_engine_AC} (b) the almost linear increase in the number of nucleated ice crystals with the increase in number of soot particles, i.e., fuel consumption. 
The number of ice crystals nucleated in ambient aerosol varies slight, 3-5\%, but remains a small portion of the total number of ice crystals.
The higher number of ice crystals is therefore due mostly to the higher number of emitted soot and partially to the larger quantity of available water vapor that allows more soot particles and ambient aerosol to form ice crystals, respectively.
The differences in terms of net radiative forcing is -22/+19\% \cref{fig:sensitivity_engine_AC} (e).

\begin{figure*}[hbt!] \centering
    \includegraphics[width=.6\textwidth]{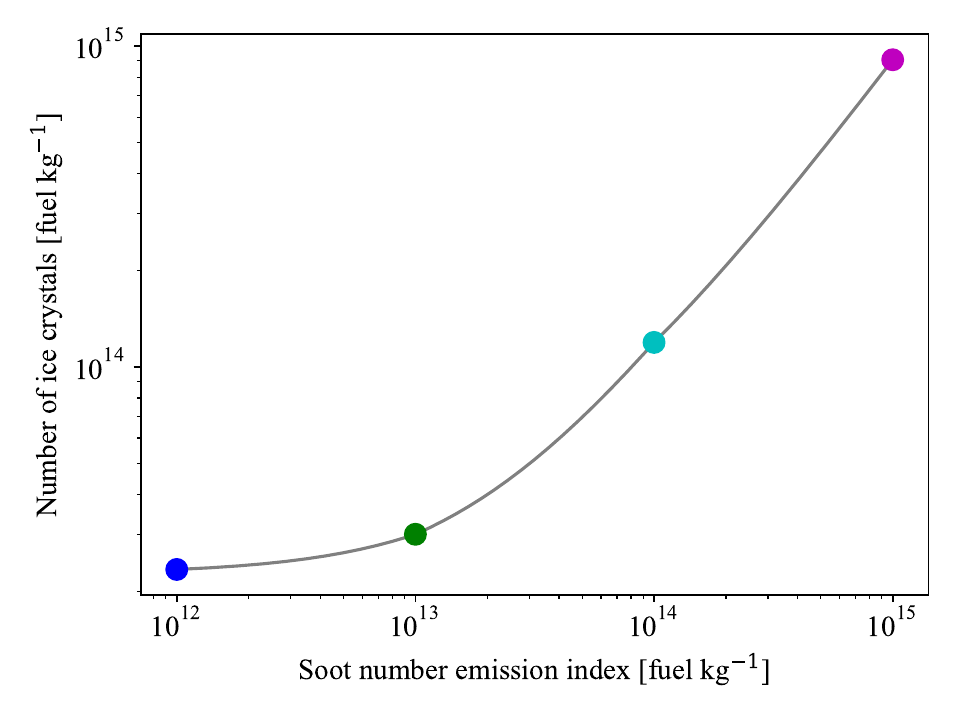}
    \caption{Sensitivity of the number of ice crystals formed on the soot number emission index per kg of fuel}
    \label{fig:EI_soot_vs_ice_crystal}
\end{figure*}

In this analysis, we also vary the order of magnitude of soot number emission index $\text{EI}_{\text{soot}}$ from the soot-poor regime (\num{e12}) to the current emission levels (\num{e15}), as identified in \cite{karcher2018formation}. 
As in all the other simulations, we do not vary the concentration of ambient aerosol.
Since the quantify of water vapor remains the same throughout this comparison (which assumes that the soot emissions would not scale with the fuel burn but instead with, e.g, the combustor characteristics, fuel type, etc.), more water vapor would be available per nucleated particle in the soot-poor regime. This is translated not only in a larger growth of the nucleated soot particles but also in more available vapor for the ambient aerosol particles to nucleate and grow.
As such, varying the soot number emission index has an effect on the activation of ice crystals both from soot particles and ambient aerosol and this effect is non-linear, cf. \cref{fig:sensitivity_engine_AC} (c) in logarithmic scale.
We can observe that the number of nucleated ice crystals on ambient aerosol more than triples between the extremes of the soot number emission index, which in turn varies by orders of magnitude. 
Toward the large emission index, $\text{EI}_{\text{soot}}=$ \num{e15} \si{\per\kilo\gram} fuel, most ice crystals form on soot particles and only a fraction on ambient aerosol. 
Toward to lower emission index, $\text{EI}_{\text{soot}}=$ \num{e12} \si{\per\kilo\gram} fuel, the ambient aerosol, represented by a higher hygroscopicity parameter, contributes to a much larger portion of the nucleated ice crystals than the soot particles.
This leads to an almost quadratic instead of linear relation between the number of formed ice crystals and the soot number emission index, as shown in \cref{fig:EI_soot_vs_ice_crystal}.
This profile resembles that of K\"{a}rcher's Fig. 3 \cite{karcher2018formation} for conditions neat the formation threshold temperature in which there is not the enhanced activation of ultrafine aqueous particles, which indeed are not presently considered in this work.

The effect on the net radiative forcing is also nonlinear, and the difference between the lower soot emissions of $\text{EI}_{\text{soot}}=$ \num{e12} $-$ \num{e13} \si{\per\kilo\gram} fuel is not as significant due to the contribution of the ambient aerosol in the lowest case.
The net radiative forcing by the highest soot emission index is still lower than the widebody aircraft case above.

\section{Conclusion} \label{sec:conclusion}

This paper details the development of a numerical framework built for high-fidelity simulations of contrails.
Our short-term goal is to simulate the initial phases of contrail formation, the jet and vortex phases, for conventional fuels and, in the long term, use particles and microphysics models suitable for contrails from alternative fuels.

In this work, we add a transport equation for water vapor, stratification, idealized jets, Lamb-Oseen vortices, and microphysics models to the charLES solver to make use of its LES Eulerian-Lagrangian capabilities for simulating the initial jet/vortex stages of contrail formation.
Our V\&V analysis (in the Appendix) agrees well with previous literature on a stratified vortex descent experiment, ice growth models, and jet phase of a contrail simulation.

Our sensitivity analysis concerns modeling particle microphysics and bypass flow, as well as a range of atmospheric, aircraft, and engine parameters.
Regarding the modeling, we find that a more complete approach to particle microphysics alters the rate of ice crystal nucleation but does not have a lasting effect on the number of ice crystals or their overall optical effect.
Adding bypass flow to the jet phase modeling also has a limited effect on the activation of ice crystals, but the addition of less saturated bypass air decreases the net radiative effect of this contrail modeling.

The sensitivity studies for the atmospheric parameters evidence a smaller sensitivity than to aircraft and engine parameters.
The aircraft size and soot number emission index have the most significant effect on the net radiative forcing.
The limited impact of the extended microphysics and the other analysis was verified for our baseline case, which is in a non-threshold condition for contrail formation. Extending this study to other extreme cases and threshold conditions for contrail formation would be helpful to obtain a more complete sensitivity analysis.

Finally, we obtain a non-linear relation between the number of soot emissions and the number of nucleated ice crystals as the ambient aerosol contributes to the ice crystal formation in the low-soot regime.
This has already been observed in previous literature, and it is a good precursor for our future studies on the impact of alternative fuels on early contrail formation.

\crefalias{section}{appendix}

\appendix

\section{V\&V of stratification and 2D vortex}  \label{app:VandV_strat}

Stratification is simulated through an initially stratified flow field and by adding source terms; for the vortex descent, we add Lamb-Oseen vortices; both of which are described in \cref{sec:IC_BC}. 
We follow the same approach and values of \cite{shirgaonkar2007large}, which are based on the original experiments \cite{sarpkaya1983trailing} and also on preceding simulations by \cite{spalart1996motion}, and we further deduce others for atmospheric conditions.
This validation case is based on the experiments of a counter-rotating vortex pair ascending in a towing tank by \cite{sarpkaya1983trailing}. These experiments were originally done in brine water to represent a stratified environment where the vortices decelerate due to the buoyancy effect.

The vortex parameters are as follows: the circulation-based Reynolds number $Re_{\Gamma}$ is 6,336, the initial vortex separation distance $b_0$ is 47.24 \si{\meter}, and the vortex descent velocity $u_{\text{vort}}$ is 1.32 \si{\meter\per\second}.
The level of stratification is defined by a Brunt-V\"{a}is\"{a}l\"{a} frequency $N_\text{BV}$ of 0.0279 \si{\per\second}, which is equivalent to the non-dimensional Brunt-V\"{a}is\"{a}l\"{a} frequency $N_\text{BV}^* = N_\text{BV}b_0/u_{\text{vort}}=1$ of the experiments. 
The core diameter is $r_c=0.25b_0$ as in Refs.~\cite{shirgaonkar2007large,spalart1996motion}.
Although this value differs from the original experiments ($r_c=0.13b_0$ \cite{sarpkaya1983trailing}), other works from literature have used other values, such as $r_c=0.2b_0$ \cite{robins1990numerical} and $r_c=0.16b_0$ \cite{delisi2000short}, yielding similar results.

To ensure that we were matching the original experiments, we also run another case with the reference values of $r_c=0.13b_0$, non-dimensional $N_\text{BV}^*$ frequency of 1, circulation-based Reynolds number $Re_{\Gamma}$ of 5,000 and a Froude number of \num{2.52e-3} (both within the experimental range).
Since the experiments were done at a small scale and in brine water, we had to make a couple of adaptations for our ideal gas solver. 
We scaled-up the initial vortex separation distance $b_0$ of about 0.103 \si{\meter} by a factor of 500 ($b_0$ about 50.7 \si{\meter}) (to speed up our explicit time-stepping simulations) and we used atmospheric conditions corresponding to an altitude of around 11,000 \si{\meter} (conditions more representative of an aircraft in the atmosphere).

\begin{figure}[hbt!]
\centering
\includegraphics[width=.5\textwidth]{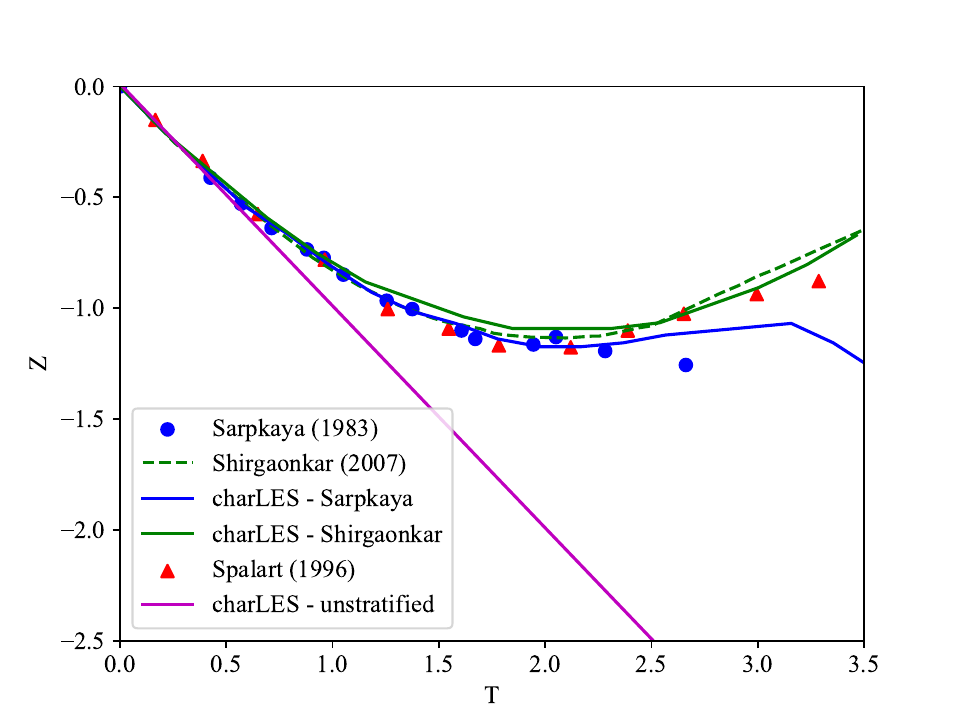}
\caption{Stratified descent of two counter-rotating vortices: comparison with experiments by \protect\cite{sarpkaya1983trailing} and simulations by \protect\cite{spalart1996motion} and \protect\cite{shirgaonkar2007large}. Simulations of both stratified and non-stratified vortex descent.}
\label{fig:stratVortDescentSarpakya1983}
\end{figure}

The vertical descent of the counter-rotating vortex pair is shown in \cref{fig:stratVortDescentSarpakya1983} following our three results labeled as \textit{charLES}: Shirgaonkar's simulation, Sarpkaya's experiment and non-stratified atmosphere. 
Time is non-dimensionalized as $T = u_{\text{vort}} t / b_0$ and the vertical location of the vortices by $Z = z / b_0$.
The results are compared with the original experiments by \cite{sarpkaya1983trailing} and simulations by \cite{spalart1996motion} and \cite{shirgaonkar2007large}.

The stratified results are reasonably close to either the simulations or the experiments, depending on the choice of vortex core diameter to vortex separation $r_c/b_0$, Froude number $Fr$, and Reynolds number $Re$. 
Slight differences may be due to the reading of the vortex pair descent, besides resolution and uncertainties in both experiments and simulations. 
We select the point of maximum vorticity to define the vortex center and follow it through the descent.

The unstratified result shows a constant velocity descent as defined by the initial circulation and, hence, equal to the initial vortex descent velocity.
This helps to visualize the relevance of the buoyancy effect induced by a strong, stable atmosphere stratification.
In this stratified environment, the vortices first descend with the initial vortex descent velocity (defined by the initial circulation), then they lose velocity as the density and pressure increase, and, in the simulations, the vortices ascend due to their buoyancy.

\section{V\&V of ice deposition model} \label{app:VandV_ice}

The ice deposition model described in \cref{sec:ice_model} is analysed against the original deposition model and a box model study.
Both reach a good agreement in terms of the particle radius and ice water content.

\paragraph*{Comparison with original model:}
In the original paper, besides presenting analytical solutions of the deposition model, \cite{karcher1996initial} also compute the ice particle radius and mass evolution. 
The authors aimed to simulate the growth of spherical, monodisperse ice particles due to water vapor supersaturation in a cooled plume.

The thermodynamic conditions are an atmospheric temperature $T_{\text{a}}$ of 223.45 \si{\kelvin}, pressure $p_{\text{a}}$ of 302.3 \si{\hecto\pascal}, and water vapor mass fraction $Y_{\text{v}_{\text{a}}}$ of \num{2.53e-4} (inferred value, corresponding to a relative humidity over ice $RH_i$ of 300\%).
The initial particle radius $r_{\text{p}}$ is 0.02 \si{\micro\meter}, the deposition coefficient $\alpha$ is equal to 0.1, and the ice number density $n_{\text{p}}$ is varied between \num{e3} \si{\per\cubic\centi\meter} and \num{e5} \si{\per\cubic\centi\meter}.

We simulate these two cases in a box model approach with no gradients until the water vapor reaches equilibrium and the particles reach their maximum radius.

We obtain a final ice water content $IWC$ close to \num{8e-2} \si{\g\per\cubic\meter} for both cases, and an averaged ice particle radius $r_{\text{p}}$ of 2.73 \si{\micro\meter} and 0.59 \si{\micro\meter} for the ice number densities $n_{\text{p}}$ of \num{e3} \si{\per\cubic\centi\meter} and \num{e5} \si{\per\cubic\centi\meter}, respectively.
This is very similar to the results of 2.75 \si{\micro\meter} and 0.6 \si{\micro\meter} of \cite{karcher1996initial}.

\begin{figure}[hbt!]
\centering
\includegraphics[width=.5\textwidth]{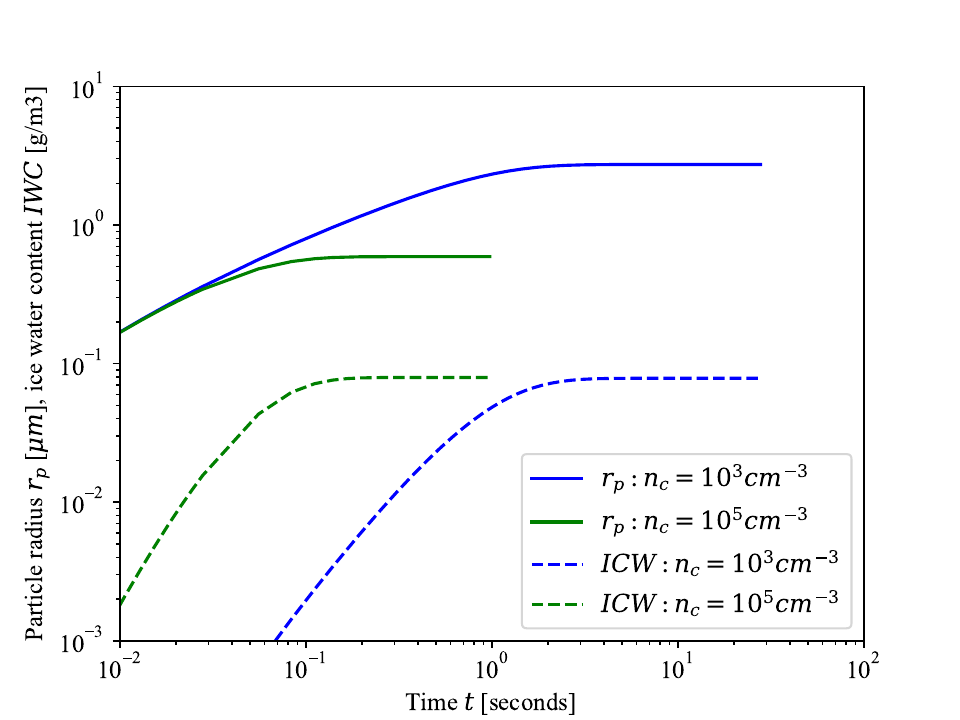}
\caption{Evolution of ice water content $ICW$ and particle radius $r_{\text{p}}$ for two levels of particle concentration following the analysis of \protect\cite{karcher1996initial}.}
\label{fig:rp_icw_karcher1996}
\end{figure}

Our results are displayed in \cref{fig:rp_icw_karcher1996}; as in Ref.~\cite{karcher1996initial}, the higher number density case depletes water vapor faster, reaching the maximum radius before the lower density scenario. 
On the other hand, the final radius is smaller since there are more particles for the same quantity of water vapor.
The final ice water content is equal for both cases since the same amount of water vapor is converted into ice, albeit at different rates.

A final note on this figure is that we produce different slopes relative to the original paper. This may be due to slight differences in the modeling, such as the growth model definition or the saturation pressure functions.
However, we do reach the same outcomes regarding equilibrium particle mass and size, which gives us confidence in our approach.

\paragraph*{Box model analysis:}
A study previously conducted by \cite{shirgaonkar2007large} in collaboration with Boeing and Aerodyne Research Inc., compared their three different microphysics models. Later, \cite{naiman2011modeling} also used this case for comparison.

To analyse the behavior of the microphysics model, simulations are run in a box model mode, as in the example above, in which the flow properties are initially homogeneous, there are no fluctuations nor gradients. The goal is to focus only on the microphysics without the influence of the flow dynamics.

The thermodynamics conditions for this case are an initial temperature $T$ and pressure $p$ of 220 \si[per-mode=symbol]{\kelvin} and 240 \si[per-mode=symbol]{\hecto\pascal}, respectively, and the relative humidity over ice $RH_i$ is set to either 110\% or 130\%. 
Particles are initially monodispersed with a radius of 0.5 \si[per-mode=symbol]{\micro\meter} and are randomly distributed in the domain with a concentration of 400 \si{\per\cubic\centi\meter}.

\begin{figure*}[hbt!] \centering
    \begin{subfigure}[t]{0.47\textwidth} \centering
        \includegraphics[width=\textwidth]{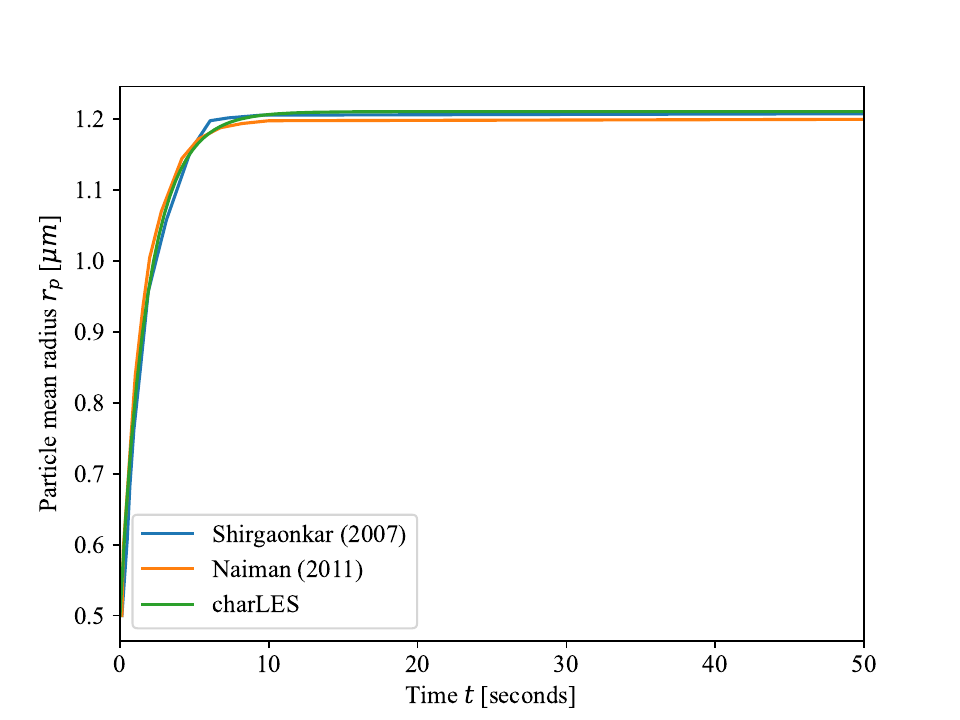}
        \caption{Evolution of mean particle radius.}
        \label{fig:val_micro_rp110}
    \end{subfigure}
\hfill
    \begin{subfigure}[t]{0.47\textwidth} \centering
        \includegraphics[width=\textwidth]{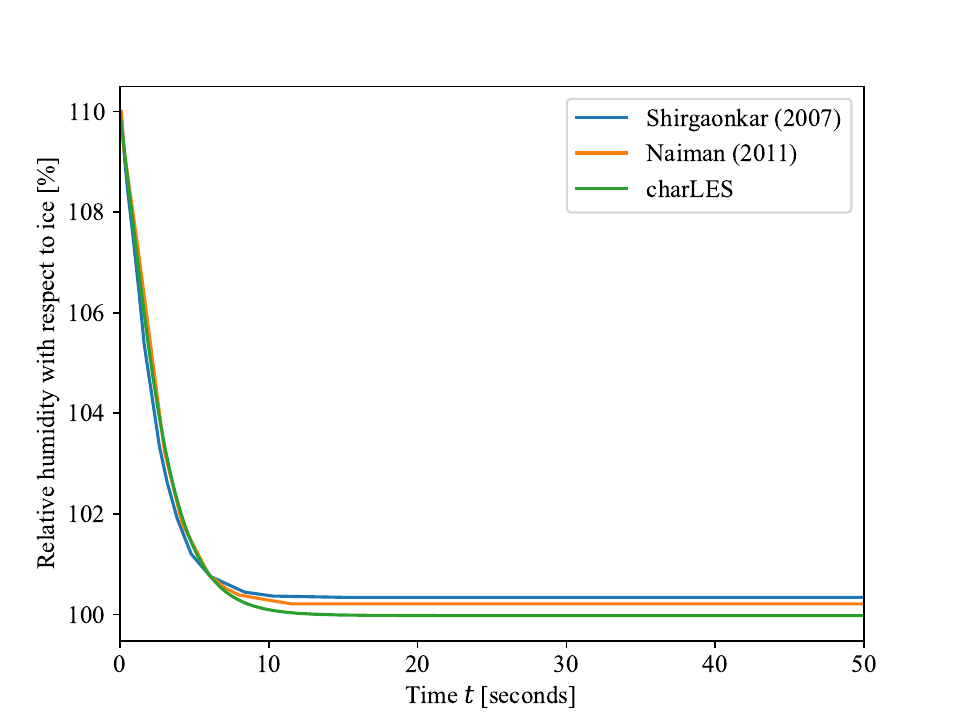}
        \caption{Evolution of mean relative humidity with respect to ice.}
        \label{fig:val_micro_rhi110}
    \end{subfigure}
\caption{Box model analysis of ice particle growth by deposition for an initial relative humidity with respect to ice of 110\%: comparison with results from \protect\cite{shirgaonkar2007large,naiman2011modeling}.}
\label{fig:val_micro_naiman}
\end{figure*}

Using the same definition for $dr/dt$, the deposition factor $\alpha$ is set equal to 1.0 and we assume that the researchers did not apply the Kelvin factor correction to the partial pressure.
We are able to match the evolution of the mean particle radius and the relative humidity over ice as shown in \cref{fig:val_micro_naiman}.
We obtain the same rate of growth and vapor depletion, which confirms that our deposition model and simulation framework behaves similarly to the one of these previous researchers simulating contrails.
We only present results for $RH_i=$110\%, but a similar match is obtained for $RH_i=$130\%.
The variation of the water vapor and ice mass are also in good agreement (not shown).

\section{Reproducing the jet phase of a contrail simulation} \label{app:VandV_jet}

In this section, we combine all the models presented in this paper, except for the stratification and idealized vortices, to represent the jet phase of a contrail.
Our aim is to verify our implementation by reproducing the results from the early simulations of \cite{paoli2004contrail} of the jet (and vortex, not compared) phase.
We have focused on the jet phase, but an initial test introducing the vortex produced reasonable results.
We chose this study due to its relative simplicity (concerning later simulations) and closeness to our numerical tools (LES, compressible solver, transport equation for water vapor).

The case we are simulating is a jet laden with particles and water vapor.
The jet is prescribed with a radial distribution of axial velocity, temperature, and water vapor, representing that of a hot jet exhaust into the atmosphere. 
The temperature of the jet is initially high, but it rapidly cools off, mixing with the atmosphere and allowing the particles to grow in excess of water vapor.

We provide a brief overview of this test case, but more details can be found in Ref.~\cite{paoli2004contrail}.
Simulations are split in two phases: the jet phase and the jet/vortex interaction phase (vortex phase).
The jet phase simulations start only with the prescribed jet and particles.
The jet is defined by a temperature $T_{\text{j}}$ of 440 \si{\kelvin}, an axial velocity $u_{x_{\text{j}}}$ of 60 \si{\meter\per\second}, and a water vapor mass fraction $Y_{\text{v}_{\text{j}}}$ of 0.3. 
The atmosphere is set to a temperature $T_{\text{a}}$ of 220 \si{\kelvin}, with no axial velocity and no background water vapor, $u_{x_{\text{a}}}=0$ \si{\meter\per\second} and $Y_{\text{v}_{\text{a}}}=0$, respectively.
The atmospheric pressure $p_{\text{a}}$ is 240 \si{\hecto\pascal} and we assume an instantaneous exhaust into the atmosphere so that $p_{\text{j}}=p_{\text{a}}=240$ \si{\hecto\pascal}. 
These values are listed in \cref{tab:jet_paoli}.

\begin{table}[hbt!] \centering
\caption{Thermodynamic conditions for the jet and atmosphere.}
\label{tab:jet_paoli} 
\begin{tabular}{l|cccc}
\hline
Properties & $T$ {[}\si{\kelvin}{]} & $p$ {[}\si{\hecto\pascal}{]}  & $u_x$ {[}\si{\meter\per\second}{]} & $Y_{\text{v}}$ \\
Atmosphere        & 220       & 240 & 0    & 0    \\
Jet        & 440       & 240 & 60   & 0.3  \\
\hline
\end{tabular}
\end{table}

Monodisperse particles are placed within the jet with a uniform distribution and a radius $r_{\text{p}}$ of 0.02 \si{\micro\meter}. 
We follow the same approach of injecting \num{2.5e5} computational particles, $n_c$, representing \num{e6} physical particles $n_{\text{p}}$ with the same size and properties.
This leads to a total of \num{2.5e11} of representative particles $n_r$, which is more coherent with engine soot number emissions.

The boundary conditions for this case are periodicity in the axial direction of the jet, and non-reflective outlet constant pressure for the remaining directions.
We have used both Cartesian and Voronoi meshes. Since our spatial discretization scheme is 2\textsuperscript{nd} order central, four orders lower than the 6\textsuperscript{th} order compact difference scheme \cite{paoli2004contrail}, we refined our mesh further from the original grid spacing $dx=dy=dz$ of 0.1 \si{\meter} to about 0.03$-$0.06 \si{\meter}.

For this case, we altered our model to be closer to the reference, e.g., using their saturation pressure equation, removing the saturation flag condition for crossing saturation pressure over water, among others.
We also set the turbulent Prandtl and Schmidt numbers, $\text{Pr}_\text{t}$ and $\text{Sc}_\text{t}$, to constant and equal to 0.3, for the rates of turbulent heat and vapor diffusion to be equal during the jet mixing with the atmosphere.
However, other differences remain between our approaches, like the definition of water vapor mean free path $\lambda_{\text{v}}$ and water vapor diffusion $D_{\text{v}}$, heat and momentum exchanges, turbulence modeling, and numerical scheme. All of these may contribute to the slight differences in results.

\begin{figure}[hbt!]
\centering
\includegraphics[width=.5\textwidth]{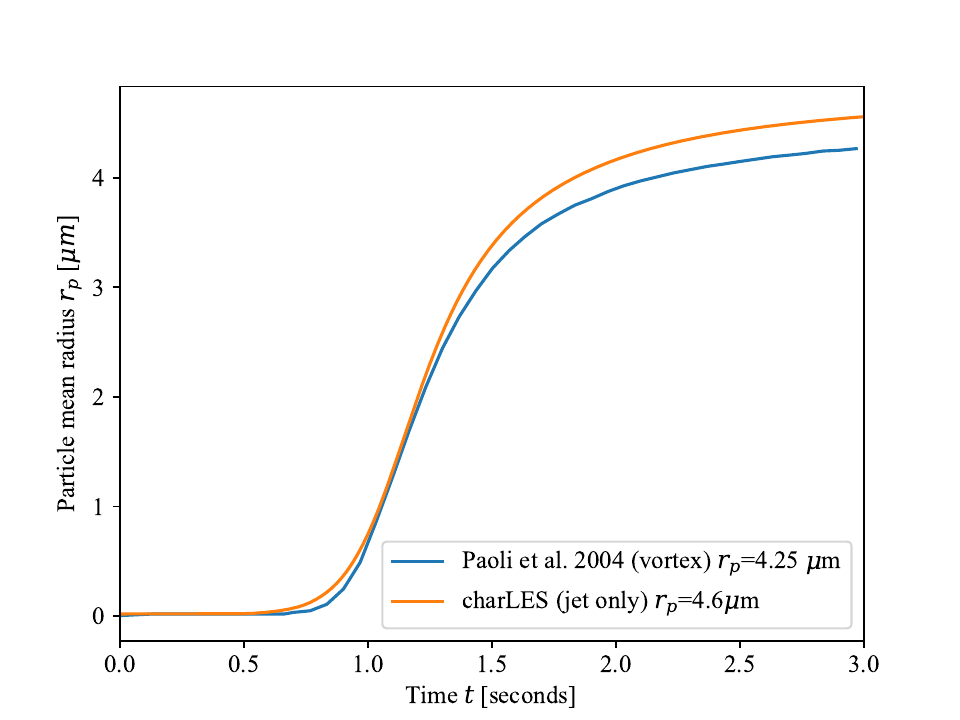}
\caption{Comparing particle mean radius of the jet phase simulation with the vortex phase of \protect\cite{paoli2004contrail}.} 
\label{fig:rp_meshM5_frontera}
\end{figure}

\begin{figure*}[hbt!] \centering
    \begin{subfigure}[t]{0.47\textwidth} \centering
        \includegraphics[width=\textwidth]{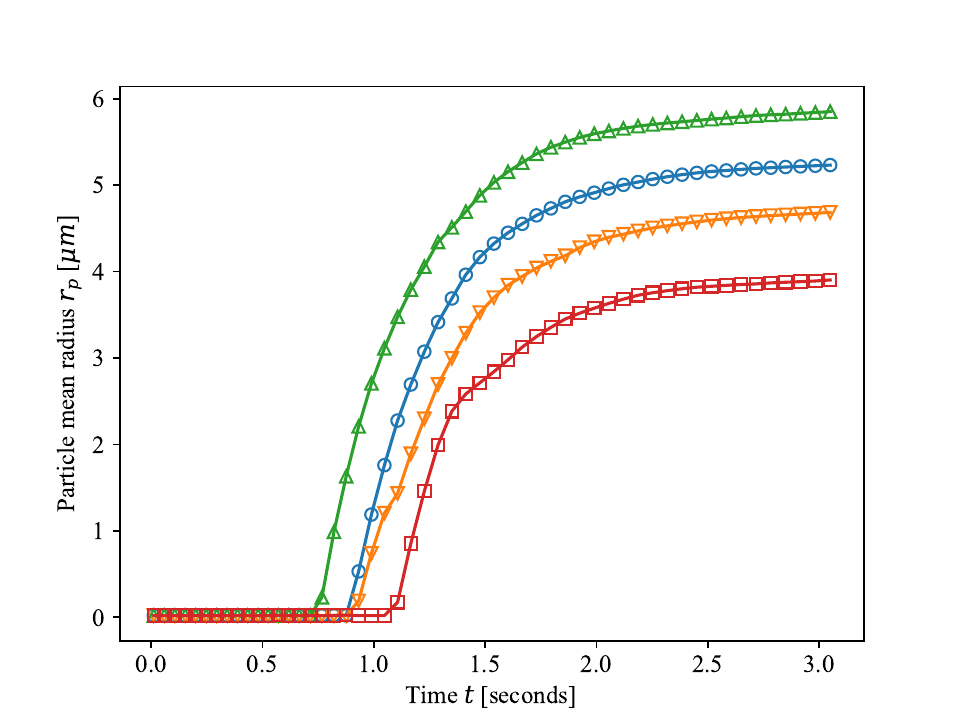}
\caption{Time history of particle radius during the jet phase.}
\label{fig:sample_rp}
    \end{subfigure}
\hfill
    \begin{subfigure}[t]{0.47\textwidth} \centering
        \includegraphics[width=\textwidth]{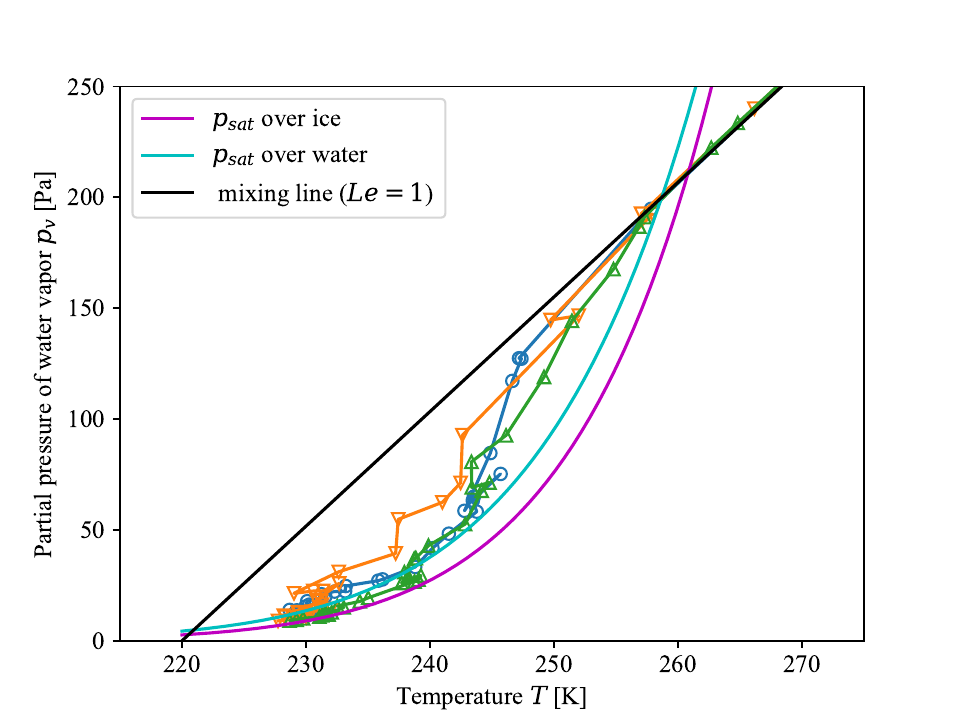}
\caption{Partial pressure vs. temperature during deposition.}
\label{fig:sample_pvT}
    \end{subfigure}
\caption{Sample of single particles to describe individual particle size growth and vapor uptake.}
\label{fig:sample_lsp}
\end{figure*}

\begin{figure}[hbt!]
\centering
\includegraphics[width=.5\textwidth]{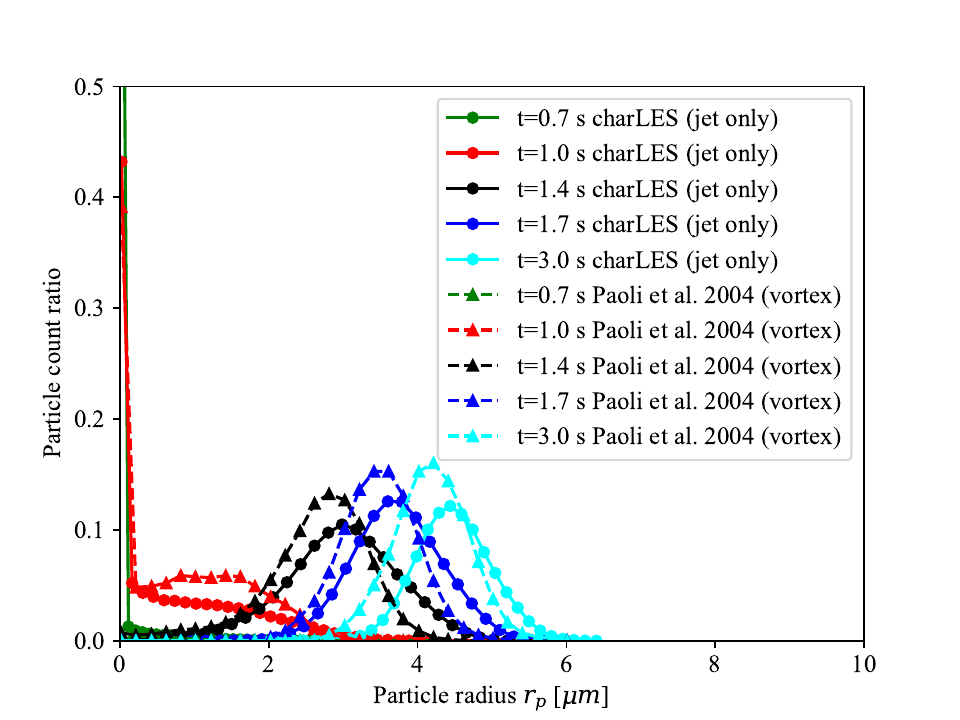}
\caption{Comparing the distribution of the particle radius during the jet only phase of our simulations and the jet/vortex interaction of \protect\cite{paoli2004contrail}.}
\label{fig:pdf_jet_PA04}
\end{figure}

Although the paper only presents the time history of the vortex phase, it mentions that the final mean particle radius is $r_{\text{p}}=4.5$ \si{\micro\meter} for the jet phase.
We can thus compare our results for the jet phase with the vortex phase of \cite{paoli2004contrail} in \cref{fig:rp_meshM5_frontera}, knowing that the plateau for the jet phase will be higher than the one presented on the plot.
In our simulations, we reach a final averaged particle radius  $r_{\text{p}}$ for the jet phase of 4.6 \si{\micro\meter}, which is very close to their value of $r_{\text{p}}=4.5$ \si{\micro\meter}. 
Our slightly larger value may be due to an earlier crossing of the saturation pressure curve, which indicates a more rapid cooling of the jet exhaust plume.

Looking into the individual particle behavior, we can observe in \cref{fig:sample_rp} the same trend as in Fig.~8 of Ref.~\cite{paoli2004contrail} where the ice particles that start deposition first grow faster, to larger radii.
Depending on the particle's location within the jet, different temperature levels, and thus, water vapor saturation, lead to the varying growth between particles.

In \cref{fig:sample_pvT}, we observe a similar evolution for the partial pressure vs. temperature around individual particles during the active growth phase, as in Fig.~8 of Ref.~\cite{paoli2002contrail}.
In this region, between the mixing line and the saturation curves, a particle, or better, the water vapor within a control volume, is supersaturated. 
As a consequence, particles can increase their mass by ice deposition and water vapor depletion, which reduces the partial pressure around the particles.
At the same time, the temperature also decreases due to the mixing processes, which explains the individual paths of the particles in the partial pressure vs temperature.
Before crossing the saturation curve (not shown), the water vapor is not supersaturated, and due to the same heat and vapor diffusive rates, the particles follow the straight mixing line.

Lastly, we compare the particle size distribution throughout the jet phase with the vortex case of \cite{paoli2004contrail} in \cref{fig:pdf_jet_PA04}.
We observe quite similar distributions, and, as expected, our case shows a tendency to larger radii as shown in the averaged particle results in \cref{fig:rp_meshM5_frontera}.

\section*{Funding Sources}

T.~S.~C.~F. acknowledges the postdoctoral fellowship funded by the Office of Naval Research from the Center for Turbulence Research at Stanford University.

\section*{Acknowledgments}

We would like to acknowledge the high-performance computing support from Cheyenne (doi:10.5065/D6RX99HX) \cite{computational2017cheyenne} provided by NCAR's Computational and Information Systems Laboratory, sponsored by the National Science Foundation, for the initial verification and validation of our contrail simulation approach.
The sensitivity analysis work used Expanse at San Diego Supercomputer Center through allocation EES230017 from the Advanced Cyberinfrastructure Coordination Ecosystem: Services \& Support (ACCESS) program, which is supported by National Science Foundation grants \#2138259, \#2138286, \#2138307, \#2137603, and \#2138296 \cite{boerner2023access}.

We are grateful for the discussions with Prof. Roberto Paoli on contrail simulation and reproducing the results of his previous paper \cite{paoli2004contrail} and with Dr. Karim Shariff and Dr. D.-G. Caprace on vortex dynamics and radiative forcing.

\bibliography{references}

\begin{thebibliography}{44}
\newcommand{\enquote}[1]{``#1''}
\providecommand{\natexlab}[1]{#1}
\providecommand{\url}[1]{\texttt{#1}}
\providecommand{\urlprefix}{URL }
\expandafter\ifx\csname urlstyle\endcsname\relax
  \providecommand{\doi}[1]{\discretionary{}{}{}https://doi.org/#1}\else
  \providecommand{\doi}[1]{\discretionary{}{}{}\urlstyle{rm}\url{https://doi.org/#1}}\fi

\bibitem[{Lee et~al.(2021)Lee, Fahey et~al.}]{lee2021contribution}
Lee, D.~S., Fahey, D.~W., et~al., \enquote{The contribution of global aviation
  to anthropogenic climate forcing for 2000 to 2018,} \emph{Atmospheric
  Environment}, Vol. 244, 2021, p. 117834.

\bibitem[{Paoli and Shariff(2016)}]{paoli2016contrail}
Paoli, R., and Shariff, K., \enquote{Contrail modeling and simulation,}
  \emph{Annual Review of Fluid Mechanics}, Vol.~48, 2016.

\bibitem[{Schumann and Heymsfield(2017)}]{schumann2017}
Schumann, U., and Heymsfield, A.~J., \enquote{On the Life Cycle of Individual
  Contrails and Contrail Cirrus,} \emph{Meteorological Monographs}, Vol.~58,
  2017, pp. 3.1 -- 3.24.
\newblock \doi{10.1175/AMSMONOGRAPHS-D-16-0005.1},
  \urlprefix\url{https://journals.ametsoc.org/view/journals/amsm/58/1/amsmonographs-d-16-0005.1.xml}.

\bibitem[{K{\"a}rcher(2018)}]{karcher2018formation}
K{\"a}rcher, B., \enquote{Formation and radiative forcing of contrail cirrus,}
  \emph{Nature communications}, Vol.~9, No.~1, 2018, p. 1824.

\bibitem[{K{\"a}rcher(1995)}]{karcher1995trajectory}
K{\"a}rcher, B., \enquote{A trajectory box model for aircraft exhaust plumes,}
  \emph{Journal of Geophysical Research: Atmospheres}, Vol. 100, No.~D9, 1995,
  pp. 18835--18844.

\bibitem[{Vancassel et~al.(2014)Vancassel, Mirabel, and
  Garnier}]{vancassel2014numerical}
Vancassel, X., Mirabel, P., and Garnier, F., \enquote{Numerical simulation of
  aerosols in an aircraft wake using a 3D LES solver and a detailed
  microphysical model,} \emph{International Journal of Sustainable Aviation},
  Vol.~1, No.~2, 2014, pp. 139--159.

\bibitem[{Lewellen(2020)}]{lewellen2020large}
Lewellen, D.~C., \enquote{A large-eddy simulation study of contrail ice number
  formation,} \emph{Journal of the Atmospheric Sciences}, Vol.~77, No.~7, 2020,
  pp. 2585--2604.

\bibitem[{Bier et~al.(2022)Bier, Unterstrasser, and Vancassel}]{bier2022box}
Bier, A., Unterstrasser, S., and Vancassel, X., \enquote{Box model trajectory
  studies of contrail formation using a particle-based cloud microphysics
  scheme,} \emph{Atmospheric Chemistry and Physics}, Vol.~22, No.~2, 2022, pp.
  823--845.

\bibitem[{Bier et~al.(2024)Bier, Unterstrasser, Zink, Hillenbrand,
  Jurkat-Witschas, and Lottermoser}]{bier2024contrail}
Bier, A., Unterstrasser, S., Zink, J., Hillenbrand, D., Jurkat-Witschas, T.,
  and Lottermoser, A., \enquote{Contrail formation on ambient aerosol particles
  for aircraft with hydrogen combustion: a box model trajectory study,}
  \emph{Atmospheric Chemistry and Physics}, Vol.~24, No.~4, 2024, pp.
  2319--2344.

\bibitem[{Paoli et~al.(2002)Paoli, H{\'e}lie, Poinsot, and
  Ghosal}]{paoli2002contrail}
Paoli, R., H{\'e}lie, J., Poinsot, T., and Ghosal, S., \enquote{Contrail
  formation in aircraft wakes using large-eddy simulations,} \emph{Studying
  Turbulence Using Numerical Simulation Databases-IX: Proceedings of the 2002
  Summer Program}, 2002.

\bibitem[{Paoli et~al.(2004)Paoli, Helie, and Poinsot}]{paoli2004contrail}
Paoli, R., Helie, J., and Poinsot, T., \enquote{Contrail formation in aircraft
  wakes,} \emph{Journal of Fluid Mechanics}, Vol. 502, 2004, pp. 361--373.

\bibitem[{Paoli et~al.(2013)Paoli, Nybelen, Picot, and
  Cariolle}]{paoli2013effects}
Paoli, R., Nybelen, L., Picot, J., and Cariolle, D., \enquote{Effects of
  jet/vortex interaction on contrail formation in supersaturated conditions,}
  \emph{Physics of Fluids}, Vol.~25, No.~5, 2013.

\bibitem[{Paugam et~al.(2010)Paugam, Paoli, and Cariolle}]{paugam2010influence}
Paugam, R., Paoli, R., and Cariolle, D., \enquote{Influence of vortex dynamics
  and atmospheric turbulence on the early evolution of a contrail,}
  \emph{Atmospheric Chemistry and Physics}, Vol.~10, No.~8, 2010, pp.
  3933--3952.

\bibitem[{Naiman et~al.(2011)Naiman, Lele, and Jacobson}]{naiman2011large}
Naiman, A., Lele, S., and Jacobson, M., \enquote{Large eddy simulations of
  contrail development: Sensitivity to initial and ambient conditions over
  first twenty minutes,} \emph{Journal of Geophysical Research: Atmospheres},
  Vol. 116, No. D21, 2011.

\bibitem[{Picot et~al.(2015)Picot, Paoli, Thouron, and
  Cariolle}]{picot2015large}
Picot, J., Paoli, R., Thouron, O., and Cariolle, D., \enquote{Large-eddy
  simulation of contrail evolution in the vortex phase and its interaction with
  atmospheric turbulence,} \emph{Atmospheric Chemistry and Physics}, Vol.~15,
  No.~13, 2015, pp. 7369--7389.

\bibitem[{Toro(2009)}]{toro2009}
Toro, E.~F., \emph{Riemann Solvers and Numerical Methods for Fluid Dynamics},
  Springer Berlin Heidelberg, 2009.
\newblock \doi{10.1007/b79761}.

\bibitem[{Vreman(2004)}]{vreman2004eddy}
Vreman, A., \enquote{An eddy-viscosity subgrid-scale model for turbulent shear
  flow: Algebraic theory and applications,} \emph{Physics of fluids}, Vol.~16,
  No.~10, 2004, pp. 3670--3681.

\bibitem[{Bier et~al.(2023)Bier, Unterstrasser, Zink, Hillenbrand,
  Jurkat-Witschas, and Lottermoser}]{bier2023contrail}
Bier, A., Unterstrasser, S., Zink, J., Hillenbrand, D., Jurkat-Witschas, T.,
  and Lottermoser, A., \enquote{Contrail formation on ambient aerosol particles
  for aircraft with hydrogen combustion: A box model trajectory study,}
  \emph{EGUsphere}, Vol. 2023, 2023, pp. 1--39.

\bibitem[{K{\"a}rcher et~al.(1996)K{\"a}rcher, Peter, Biermann, and
  Schumann}]{karcher1996initial}
K{\"a}rcher, B., Peter, T., Biermann, U.~M., and Schumann, U., \enquote{The
  initial composition of jet condensation trails,} \emph{Journal of atmospheric
  sciences}, Vol.~53, No.~21, 1996, pp. 3066--3083.

\bibitem[{Pruppacher and Klett(2010)}]{pruppacher2010microphysics}
Pruppacher, H.~R., and Klett, J.~D., \emph{Microphysics of clouds and
  precipitation}, second revised and enlarged edition with an introduction to
  cloud chemistry and cloud electricity ed., Springer Science \& Business
  Media, 2010.

\bibitem[{Petters and Kreidenweis(2007)}]{petters2007single}
Petters, M., and Kreidenweis, S., \enquote{A single parameter representation of
  hygroscopic growth and cloud condensation nucleus activity,}
  \emph{Atmospheric Chemistry and Physics}, Vol.~7, No.~8, 2007, pp.
  1961--1971.

\bibitem[{K{\"a}rcher et~al.(2015)K{\"a}rcher, Burkhardt, Bier, Bock, and
  Ford}]{karcher2015microphysical}
K{\"a}rcher, B., Burkhardt, U., Bier, A., Bock, L., and Ford, I., \enquote{The
  microphysical pathway to contrail formation,} \emph{Journal of Geophysical
  Research: Atmospheres}, Vol. 120, No.~15, 2015, pp. 7893--7927.

\bibitem[{Murphy and Koop(2005)}]{murphy2005review}
Murphy, D.~M., and Koop, T., \enquote{Review of the vapour pressures of ice and
  supercooled water for atmospheric applications,} \emph{Quarterly Journal of
  the Royal Meteorological Society: A journal of the atmospheric sciences,
  applied meteorology and physical oceanography}, Vol. 131, No. 608, 2005, pp.
  1539--1565.

\bibitem[{Garnier et~al.(1997)Garnier, Baudoin, Woods, and
  Louisnard}]{garnier1997engine}
Garnier, F., Baudoin, C., Woods, P., and Louisnard, N., \enquote{Engine
  emission alteration in the near field of an aircraft,} \emph{Atmospheric
  Environment}, Vol.~31, No.~12, 1997, pp. 1767--1781.

\bibitem[{van~de Hulst(1981)}]{hulst1981light}
van~de Hulst, H.~C., \emph{Light scattering by small particles}, Courier
  Corporation, 1981.

\bibitem[{K{\"a}rcher et~al.(2009)K{\"a}rcher, Mayer, Gierens, Burkhardt,
  Mannstein, and Chatterjee}]{karcher2009aerodynamic}
K{\"a}rcher, B., Mayer, B., Gierens, K., Burkhardt, U., Mannstein, H., and
  Chatterjee, R., \enquote{Aerodynamic contrails: Microphysics and optical
  properties,} \emph{Journal of the atmospheric sciences}, Vol.~66, No.~2,
  2009, pp. 227--243.

\bibitem[{Sanz-Mor{\`e}re et~al.(2021)Sanz-Mor{\`e}re, Eastham, Allroggen,
  Speth, and Barrett}]{sanz2021impacts}
Sanz-Mor{\`e}re, I., Eastham, S.~D., Allroggen, F., Speth, R.~L., and Barrett,
  S.~R., \enquote{Impacts of multi-layer overlap on contrail radiative
  forcing,} \emph{Atmospheric Chemistry and Physics}, Vol.~21, No.~3, 2021, pp.
  1649--1681.

\bibitem[{Corti and Peter(2009)}]{corti2009simple}
Corti, T., and Peter, T., \enquote{A simple model for cloud radiative forcing,}
  \emph{Atmospheric Chemistry and Physics}, Vol.~9, No.~15, 2009, pp.
  5751--5758.

\bibitem[{Gago et~al.(2002)Gago, Brunet, and Garnier}]{gago2002numerical}
Gago, C.~F., Brunet, S., and Garnier, F., \enquote{Numerical investigation of
  turbulent mixing in a jet/wake vortex interaction,} \emph{AIAA journal},
  Vol.~40, No.~2, 2002, pp. 276--284.

\bibitem[{Paoli et~al.(2003)Paoli, Laporte, Cuenot, and
  Poinsot}]{paoli2003dynamics}
Paoli, R., Laporte, F., Cuenot, B., and Poinsot, T., \enquote{Dynamics and
  mixing in jet/vortex interactions,} \emph{Physics of fluids}, Vol.~15, No.~7,
  2003, pp. 1843--1860.

\bibitem[{Schumann et~al.(1997)Schumann, D{\"o}rnbrack, D{\"u}rbeck, and
  Gerz}]{schumann1997large}
Schumann, U., D{\"o}rnbrack, A., D{\"u}rbeck, T., and Gerz, T.,
  \enquote{Large-eddy simulation of turbulence in the free atmosphere and
  behind aircraft,} \emph{Fluid dynamics research}, Vol.~20, No. 1-6, 1997,
  p.~1.

\bibitem[{Schumann et~al.(1998)Schumann, Schlager, Arnold, Baumann,
  Haschberger, and Klemm}]{schumann1998dilution}
Schumann, U., Schlager, H., Arnold, F., Baumann, R., Haschberger, P., and
  Klemm, O., \enquote{Dilution of aircraft exhaust plumes at cruise altitudes,}
  \emph{Atmospheric Environment}, Vol.~32, No.~18, 1998, pp. 3097--3103.

\bibitem[{Cantin et~al.(2022)Cantin, Chouak, Morency, and
  Garnier}]{cantin2022eulerian}
Cantin, S., Chouak, M., Morency, F., and Garnier, F.,
  \enquote{Eulerian--Lagrangian CFD-microphysics modeling of a near-field
  contrail from a realistic turbofan,} \emph{International Journal of Engine
  Research}, Vol.~23, No.~4, 2022, pp. 661--677.

\bibitem[{Khou et~al.(2015)Khou, Ghedhaifi, Vancassel, and
  Garnier}]{khou2015spatial}
Khou, J.-C., Ghedhaifi, W., Vancassel, X., and Garnier, F., \enquote{Spatial
  simulation of contrail formation in near-field of commercial aircraft,}
  \emph{Journal of Aircraft}, Vol.~52, No.~6, 2015, pp. 1927--1938.

\bibitem[{Khou et~al.(2017)Khou, Ghedha{\"\i}fi, Vancassel, Montreuil, and
  Garnier}]{khou2017cfd}
Khou, J., Ghedha{\"\i}fi, W., Vancassel, X., Montreuil, E., and Garnier, F.,
  \enquote{CFD simulation of contrail formation in the near field of a
  commercial aircraft: Effect of fuel sulfur content,} \emph{Meteorologische
  Zeitschrift}, Vol.~26, No.~6, 2017, pp. 585--596.

\bibitem[{Ferreira et~al.(2024)Ferreira, Alonso, and
  Gorl{\'e}}]{ferreira2024developing}
Ferreira, T., Alonso, J.~J., and Gorl{\'e}, C., \enquote{Developing a Numerical
  Framework for the High-Fidelity Simulation of Contrails: Sensitivity Analysis
  for Conventional Contrails,} \emph{AIAA AVIATION FORUM AND ASCEND 2024},
  2024, p. 3775.

\bibitem[{Shirgaonkar and Lele(2007)}]{shirgaonkar2007large}
Shirgaonkar, A.~A., and Lele, S.~K., \enquote{Large eddy simulation of early
  stage aircraft contrails,} Report No. TF-100, Stanford University, 2007.

\bibitem[{Sarpkaya(1983)}]{sarpkaya1983trailing}
Sarpkaya, T., \enquote{Trailing vortices in homogeneous and density-stratified
  media,} \emph{Journal of Fluid Mechanics}, Vol. 136, 1983, pp. 85--109.

\bibitem[{Spalart(1996)}]{spalart1996motion}
Spalart, P.~R., \enquote{On the motion of laminar wing wakes in a stratified
  fluid,} \emph{Journal of Fluid Mechanics}, Vol. 327, 1996, pp. 139--160.

\bibitem[{Robins and Delisi(1990)}]{robins1990numerical}
Robins, R.~E., and Delisi, D.~P., \enquote{Numerical study of vertical shear
  and stratification effects on the evolution of a vortex pair,} \emph{AIAA
  journal}, Vol.~28, No.~4, 1990, pp. 661--669.

\bibitem[{Delisi and Robins(2000)}]{delisi2000short}
Delisi, D.~P., and Robins, R.~E., \enquote{Short-scale instabilities in
  trailing wake vortices in a stratified fluid,} \emph{AIAA journal}, Vol.~38,
  No.~10, 2000, pp. 1916--1923.

\bibitem[{Naiman(2011)}]{naiman2011modeling}
Naiman, A.~D., \enquote{Modeling aircraft contrails and emission plumes for
  climate impacts,} Ph.D. thesis, Stanford University, 2011.

\bibitem[{Computational and Laboratory(2017)}]{computational2017cheyenne}
Computational, and Laboratory, I.~S., \enquote{Cheyenne: HPE/SGI ICE XA system
  (university community computing),} , 2017.
\newblock \doi{10.5065/D6RX99HX}.

\bibitem[{Boerner et~al.(2023)Boerner, Deems, Furlani, Knuth, and
  Towns}]{boerner2023access}
Boerner, T.~J., Deems, S., Furlani, T.~R., Knuth, S.~L., and Towns, J.,
  \enquote{{ACCESS: Advancing innovation: NSF’s advanced cyberinfrastructure
  coordination ecosystem: Services \& support},} \emph{Practice and Experience
  in Advanced Research Computing}, NA, 2023, pp. 173--176.

\end{thebibliography}

\end{document}